\documentclass[aps,twocolumn,10pt,
superscriptaddress,
footinbib,
noeprint,
longbibliography,
prb]{revtex4-2}

\usepackage{graphicx}
\usepackage{dcolumn}
\usepackage{bm}
\usepackage{graphics}
\usepackage{epsfig,color}
\usepackage{physics2}
\usephysicsmodule{braket}
\usepackage{dsfont}
\usepackage{mathtools,amssymb}
\usepackage{multirow}

\usepackage[breaklinks,colorlinks,bookmarks=false,citecolor=red,linkcolor=blue,urlcolor=blue]{hyperref}

\begin{document} 

\title{Cluster truncated Wigner approximation for bond-disordered Heisenberg spin models}

\author{Adrian Braemer}
\email[]{adrian.braemer@physi.uni-heidelberg.de}
\affiliation{Physikalisches Institut, Universit\"at Heidelberg, Im Neuenheimer Feld 226, 69120 Heidelberg, Germany}

\author{Javad Vahedi}
\email[]{javad.vahedi@uni-jena.de }
\affiliation{Institute of Condensed Matter Theory and Optics, Friedrich-Schiller-University Jena, Max-Wien-Platz 1, 07743 Jena, Germany}

\author{Martin Gärttner}
\email[]{martin.gaerttner@uni-jena.de}
\affiliation{Institute of Condensed Matter Theory and Optics, Friedrich-Schiller-University Jena, Max-Wien-Platz 1, 07743 Jena, Germany}

\date{\today}
\begin{abstract}
We present a comprehensive numerical investigation of the cluster Truncated Wigner Approximation (cTWA) applied to quench dynamics in bond-disordered Heisenberg spin chains with power-law interactions. 
We find that cTWA yields highly accurate results over a wide parameter range. However, its accuracy hinges on a suitable choice of clusters. By using a clustering strategy inspired by the strong disorder renormalisation group (SDRG)/real-space renormalization group (RSRG), clusters of two spins are sufficient to obtain essentially exact results in the regime of strong disorder. Surprisingly, even for rather weak disorder, e.g.\ in the presence of very long-range interactions, this choice of clustering outperforms a naive choice of clusters of consecutive spins. Additionally, we develop a discrete sampling scheme for the initial Wigner function, as an alternative to the originally introduced scheme based on Gaussian approximations. This sampling scheme puts cTWA on the same conceptional footing as regular dTWA for single spins and yields some reduction in the Monte Carlo shot noise compared to the Gaussian scheme. 
\end{abstract}
\maketitle

\section{Introduction}\label{sec1}
Long-range interactions arise in several physical scenarios within disordered quantum many-body systems. For example, in doped semiconductors containing randomly positioned magnetic impurities, interactions occur via exchange couplings that depend on their spatial separation~\cite{Anderson1958, MOTT1976, KETTEMANN2023}. These interactions exhibit different behaviors depending on the state of the system. In insulating phases, the interaction strength decreases exponentially, as denoted by $J(r)\propto \exp{(-r/\xi)}$, while in metallic phases the interactions operate through the RKKY mechanism, following a power-law decay described by $J(r)\propto r^{-d}$, where $d$ represents the dimension of the host system. Interestingly, sufficiently random, power-law interacting systems can even feature ultra-slow relaxation known from classical spin glasses as observed in local two-level systems formed by tunneling ions interacting through dipole-dipole and elastic forces~\cite{Cesare1990,Cesare1992}.

Moreover, recent experimental progress has enabled the manipulation and investigation of cold atoms or molecules featuring strong dipole-dipole interactions in diverse setups, including optical lattices~\cite{Sharma2013,Yan2013,Fersterer2019}, Rydberg gases~\cite{Zeiher2017,Bernien2017,Keesling2019}, and trapped ions~\cite{Islam2013,Jurcevic2014,Richerme2014,Grttner2017,Whitlock2021}. This has, in turn, spurred theoretical interest in studying quantum many-body dynamics in systems characterized by varying interaction ranges.

However, the potential of these studies is often limited by the lack of suitable computational tools. Considering that the Hilbert space of the system grows exponentially with the system size, the exact solution of quantum dynamics is limited to rather small systems. Even employing sophisticated tools, e.g.\ based on Krylov subspaces \cite{Nauts1983,Park1986,Colmenarez2020,vahedi2022,Faridfar2022}, typically allows simulating systems of a only few tens of spins. Leaving the realm of exact methods, one usually tries to approximate the wave functions with a variational ansatz such as matrix product states (MPS)~\cite{schollwoeckDensitymatrixRenormalizationGroup2011} and solves the dynamics within this variational manifold. While these MPS based methods, such as time-dependent density matrix renormalization group, have been used very successfully to simulate large, one-dimensional many-body systems with nearest neighbor interactions~\cite{Vidal2003,White2004}, they struggle for higher dimensional or long-range interacting systems due to the rapid generation of entanglement~\cite{Zaletel2015, Schollwock2019}.

In the search for effective approaches to deal with many-body systems and the entanglement problem, phase-space methods have emerged as promising candidates. Among them, the truncated Wigner approximation (TWA)~\cite{Blakie2008,Polkovnikov2010}, based on the Wigner-Weyl correspondence, stands out as a practical and widely adaptable strategy for exploring the dynamics of quantum many-body systems, even in higher-dimensional settings with long-range interactions~\cite{Tuchman2006,Davidson2017,Nagao2019,muleady2023validating}. At its heart, TWA approximates the dynamics of the Wigner function, i.e.\ the phase space analogue of the wave function, by particles following the classical mean-field equations of motions. The initial conditions of these particles are sampled from a Gaussian approximation of the initial Wigner function. While a priori TWA is exact only for short times, numerical experiments have shown it to yield accurate results at intermediate or even late times in some cases~\cite{minkHybridDiscretecontinuousTruncated2022}. 

Although TWA was originally developed in the context of bosonic systems where a clear classical limit exists, it can also be applied to spin systems. Remarkably, for finite-dimensional quantum systems there exists a discrete formulation of the quantum phase-space~\cite{WOOTTERS19871}. For spin systems prepared in a product state, discrete TWA (dTWA)  exploits this to dramatically improve accuracy~\cite{Schachenmayer2015} and to capture quantum revivals ~\cite{Pikovski2015,Pucci2016,Acevedo2017,Arghavan2017,Czischek2018,Berges2018,Bhuvanesh2019,Khasseh2020,Whitlock2021,Masaya2021,Morong2021}.

Another extension aims at incorporating more quantum interactions into the equations of motion, which in traditional TWA are essentially mean-field equations for single particles~\cite{Davidson2015}. This so-called cluster TWA (cTWA) does so by grouping spins together into clusters and then assigning classical variables to all degrees of freedom within these clusters~\cite{Wurtz2018}. Thus all quantum interaction within a cluster a treated exactly and only interactions between clusters are approximated semi-classically. In the limit of clusters consisting of single spins, cTWA is identical to (d)TWA while in the opposing limit, where the whole system constitutes a single cluster the exact quantum evolution is recovered. Thus one has a tuning parameter to assess the convergence of the simulation which the usual (d)TWA lacks. While in principle cTWA is compatible with the discrete phase space formulation, literature on their combination is quite sparse. A conceptual precursor, dubbed GDTWA, exists in \cite{Zhu2019} where the discrete sampling was extended to larger $SU(N)$ spins. In a recent preprint a variant of discrete sampling is applied to a Bose-Hubbard model~\cite{nagaoTwodimensionalCorrelationPropagation2024}.

In this paper, we present a generalization of both cTWA and dTWA combining the discrete sampling scheme of the latter with the capability of treating clusters of spins of the former, which we term dcTWA. We then systematically evaluate the performance of these methods in the context of quench dynamics for bond disordered XX and XXZ long-range interacting spin $1/2$ models. More precisely, we study the dynamics of an initial N\'eel state by means of the decay of the staggered magnetization and the buildup of R\'enyi entropy in a two-spin subsystem for different interaction ranges and disorder strengths and compare the results from the semi-classical methods to exact diagonalization. While in the weakly disordered regime a bigger cluster size is beneficial generally, we find that in the strongly disordered regime the physics is well captured by clusters of size 2 if they are chosen following a pairing rule known from the real space renormalization group (RSRG) approach to bond-disordered systems. Our analysis of the statistical uncertainties reveals that although the averaged results from cTWA and dcTWA are similar, dcTWA shows less sampling noise and thus converges faster.

\begin{figure}
    \includegraphics[width=1. \columnwidth]{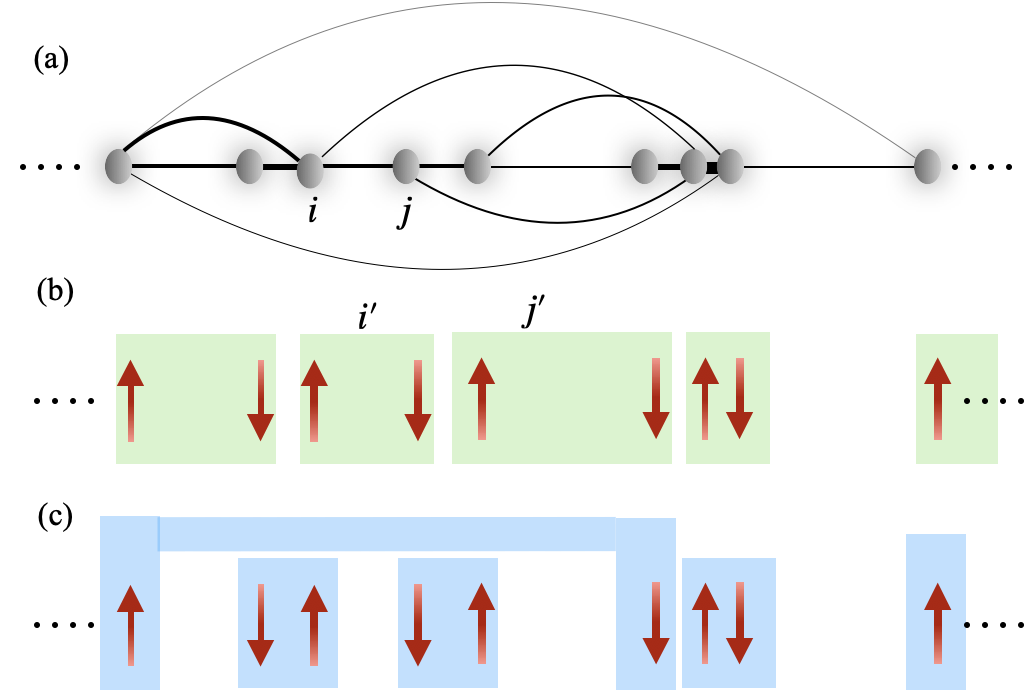}
    \caption{ (a) This diagram shows a long-range bond disorder spin chain where spins are randomly positioned along a lattice. The couplings are represented by solid lines, $J_{ij}$, while stronger bonds are indicated by thicker lines. (b) The initial state is the N\'eel state, represented as $\ket{\Psi_0} = \ket{\uparrow\downarrow\cdots\uparrow\downarrow}$, where each arrow represents the spin direction. The green shaded box illustrates a naive clustering of spins into clusters of size-2. (c) Clustering inspired by the real space renormalization group, as explained in the text. }
    \label{fig1}
\end{figure}

\section{MODEL AND METHODS}
\subsection{Model}
We study the behavior of a disordered spin chain with long-range interactions, described by the Hamiltonian 

\begin{equation}
H=\sum_{i<j} J_{ij} (\hat{s}_i^x \hat{s}_j^x + \hat{s}_i^y \hat{s}_j^y + \Delta \hat{s}_i^z \hat{s}_j^z)
\end{equation}
where $N$ spins ($\hat{s}_i=\frac{1}{2}\hat{\sigma}_i$) are randomly positioned at locations $r_i$ along a lattice of length $L$ with lattice spacing $a$, resulting in a density $f=N/L$. The interactions $J_{ij}$ between pairs of sites $i$ and $j$ are long-range, characterised by a power-law decay with parameter $\alpha$:  $J_{ij} = J_0 |(r_i - r_j)/a|^{-\alpha}$. Throughout our study we set $J_0 = 1$ and $a = 1$, and employ open boundary conditions.

The disorder in this model arises from the random arrangement of spins along the chain, leading to different spin-spin couplings. The ground-state entanglement properties of this system (with $\Delta=0$) have been previously studied in Ref.~\cite{Mohdeb2020}, where it was found that the entanglement entropy (EE) exhibits a logarithmic enhancement at zero temperature, independent of the parameter $\alpha$. Moreover, it was found that for $\alpha>\alpha^*$ the EE of the excited eigenstates reflects the logarithmic divergence observed in the ground state. Conversely, for $\alpha<\alpha^*$, the EE exhibits an algebraic growth with respect to the subsystem size $l$, characterised by $s_l\sim l^\beta$, where $0<\beta < 1$~\cite{Mohdeb2022, Mohdeb2023}.

We explore the system dynamics by initializing it in a N\'eel state and subsequently computing dynamic observables. These observables encompass the staggered magnetization $M^{\rm st}(t) = \sum_i (-1)^i \langle \hat{\sigma}_i^z(t) \rangle / N$ and the R\'enyi-2 entropy $S_2(t)$ evaluated over a two-spin subsystem. The R\'enyi-2 entropy belongs to a continuum of entropy measures defined as $S_\gamma(\hat\rho_A(t)) = \frac{1}{1 - \gamma} \log_2 \left( \text{tr}{[\hat\rho_A(t)^\gamma]} \right)$, where $\gamma \ge 1$. In this context, $\hat\rho_A(t) \equiv \text{Tr}_B \hat\rho(t)$ signifies the reduced density matrix associated with a subsystem $A$, and $\hat\rho(t)$ represents the density matrix of the entire system. Expanding the two-site reduced density matrix $\hat{\rho}_{ij}=\frac{1}{4}\sum_{\alpha\beta}\langle\hat{\sigma}^\alpha_i\hat{\sigma}^\beta_j\rangle\hat{\sigma}^\alpha_i\hat{\sigma}^\beta_j$ in a basis of Pauli strings gives a clear recipe for extracting the R\'enyi-2 entropy from the expectation values of observables:
\begin{align}
    S_2(\hat\rho_{ij}(t)) &= -\log_2\left(\mathrm{Tr}\left[\hat{\rho}_{ij}(t)\right]^2\right)\\
    &= 2 - \log_2 \left( \sum_{\alpha\beta} \left\langle\hat{\sigma}^\alpha_i\hat{\sigma}^\beta_j\right\rangle^2\right)\label{eq:TWA-renyi2}
\end{align}
where we used the trace orthogonality of the Pauli strings. This expression has a clear physical meaning: The more correlations the subsystem retains after tracing out the environment, the weaker the entanglement is.

\subsection{Cluster Truncated Wigner Approximation (cTWA)}
Phase-space methods are powerful tools for simulating quantum system dynamics close to the classical limit. These methods have applications across various scientific domains, including quantum chemistry, optics, and condensed matter physics~\cite{WOOTTERS19871,Polkovnikov2010}. Among them, the TWA maps quantum degrees of freedom onto classical phase-space variables following classical equations of motion as in a mean-field treatment. Quantum fluctuations are taken into account by Monte Carlo sampling of initial conditions from the Wigner function, which guarantees accuracy on short time scales. However, for quantum systems close to the classical limit, e.g.\ highly occupied bosonic modes or collective spin models, TWA has been found to yield accurate results even at late times~\cite{Polkovnikov2010}.

When applying TWA to spin systems, usually one considers 3 degrees of freedom per spin: Its $X$, $Y$ and $Z$ magnetization~\cite{Schachenmayer2015}. Mapping these to classical variables treats all quantum interactions between spins on a mean-field level, which is justified if the interactions are either weak or very long-range and thus average out~\cite{minkHybridDiscretecontinuousTruncated2022}. One avenue of incorporating more quantum effects into the dynamics, known as cluster TWA (cTWA), uses the degrees of freedom of clusters of spins instead of just the single spin ones~\cite{Wurtz2018}. In effect, this means all quantum dynamics within a cluster is computed exactly and only the interaction between clusters is approximated on a mean-field level. In the limit of a single cluster encompassing the whole system, cTWA is equivalent to an exact solution. Conversely, in the limit of clusters of single spins, cTWA reduces to regular TWA. Thus cTWA offers a tuning parameter to steadily tune between TWA and an exact solution by means of increasing the cluster size. In order to be self-contained, we provide an overview of this method. For a more detailed introduction, we refer the reader to the paper by Wurtz~et~al.~\cite{Wurtz2018}.

\begin{figure}
    \centering
    \includegraphics[width=0.9\columnwidth]{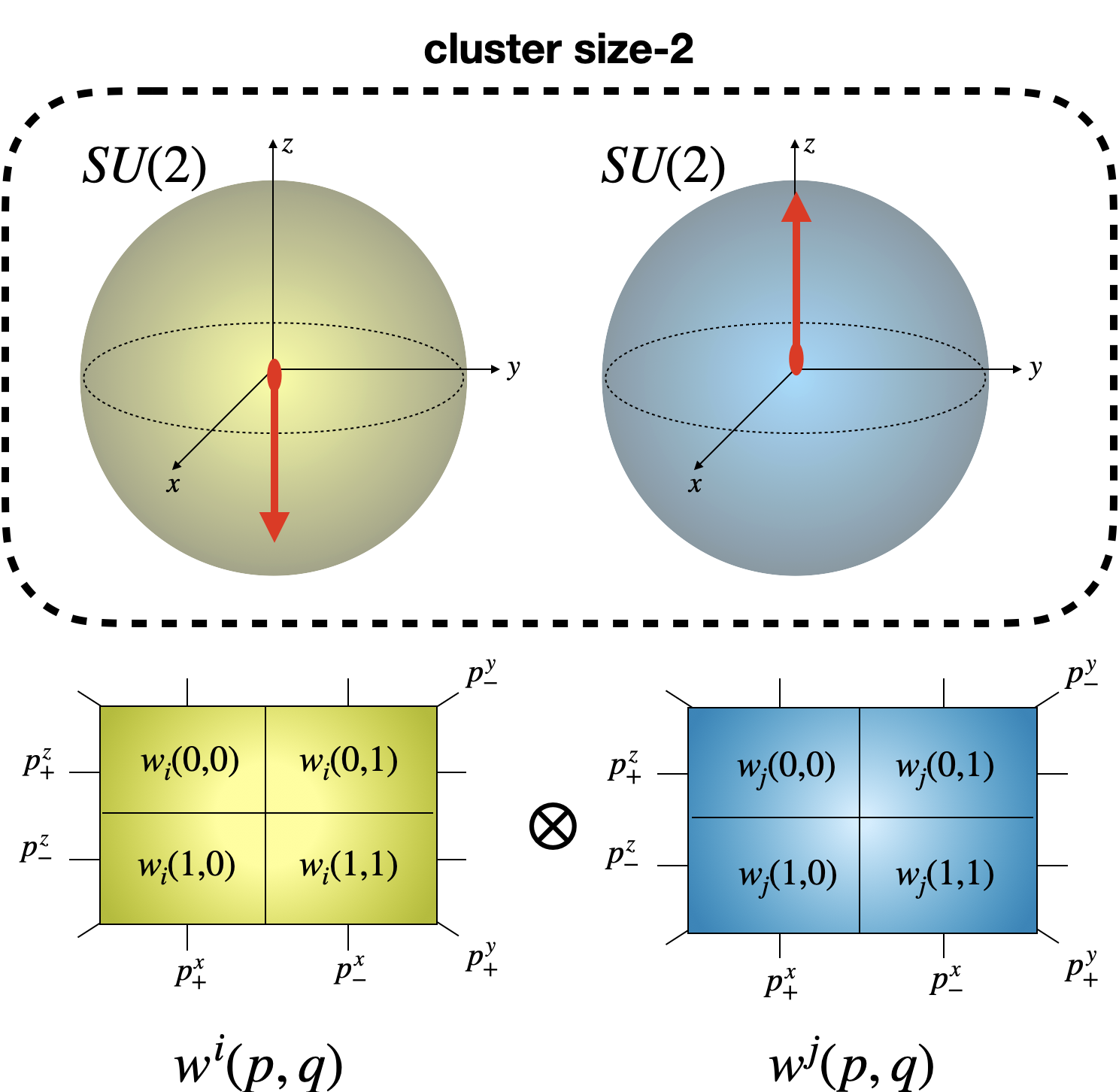}
    \caption{A schematic of the discrete cluster truncated Wigner approximation. Considering a cluster of two spins, the individual Hilbert spaces (depicted as Bloch spheres) combine to the Hilbert space of the cluster. Shown below is a representation of the single spin discrete Wigner functions in the spirit of~\cite{Schachenmayer2015}. The probabilities of a spin pointing along the $\pm x$, $\pm y$ and $\pm z$ directions are computed by summing over the vertical, diagonal and horizontal lines, respectively. For product states within a cluster, one can simply take the tensor product of single spin discrete Wigner functions to obtain Wigner functions for the cluster. In the case of clusters of size 2, the resulting Wigner function is four-dimensional and contains 16 phase-points.}
    \label{fig2}
\end{figure}

To illustrate the cTWA method, consider a system of interacting spins-$1/2$ described by the Hamiltonian:

\begin{equation}
H=\sum_{ij}J_{ab}^{ij}\hat\sigma_a^i\hat\sigma_b^j+\sum_jB^j_a\hat\sigma_a^j
\end{equation}
Here, $a, b \in x,y,z$ are the indices of Pauli matrices, and $i, j$ denote distinct spins on the lattice. The couplings $J_{ab}^{ij}$ and fields $B^j_a$ can be either short-range or long-range.

The following steps outline the implementation of the operator cTWA:
\begin{itemize}
    
    \item[$\circ$] Divide the system into clusters indexed by $[i']$, as shown in Fig.~\ref{fig1}(b). Define a complete operator basis $\{\hat{X}_p^{[i']}\}, p=0,\cdots,D^2-1$ for the Hilbert space of each cluster, where $D=2^n$ is the dimension of the Hilbert space and $n$ the number of spins making up the cluster. Ensure that the basis operators are trace-orthogonal and satisfy ${\rm Tr}[\hat X_p^{[i']}\hat{X}_q^{[j']}]=D\delta_{pq}\delta^{[i'][j']}$. Then any operator $O^{[i']}$ inside a cluster $[i']$ can be written as linear combination of the basis operators $\hat{O}^{[i']}=\sum_p o_p \hat{X}^{[i']}_p$.
    
    \item[$\circ$]  Define structure constants $f_{pqr}$ as
    \begin{equation}
        [\hat{X}^{[i']}_p,\hat{X}^{[j']}_q]=if_{pqr}\delta_{[i'][j']} \hat{X}^{[i']}_r,
    \end{equation}
    which project commutators onto the basis spanned by $\{\hat{X}^{[i']}_p\}$.

    \item[$\circ$] Express the Hamiltonian in terms of cluster operators $\hat X^{i'}_\alpha$.
    \begin{equation}
    \hat{H}=\sum_{[i'][j']}\mathbf{J}_{pq}^{[i'][j']}\hat{X}^{[i']}_p\hat{X}^{[j']}_q+\sum_{[j']}\mathbf{B}^{[j']}_p\hat{X}^{[j']}_p
    \end{equation}
    The interactions $\mathbf{J}$ and fields $\mathbf{B}$ generally differ from the original parameters $J_{ab}^{ij}$, $B^j_a$. For instance, local fields now encompass connections among spins residing within a particular cluster, given that an operator $\sigma_a^i\sigma_p^j$ becomes linear in $\hat{X}^{[j']}_p$ when both spins $i$ and $j$ are part of the same cluster $[j']$.

    \item[$\circ$] Associate basis operators $\hat{X}^{[i']}_p$ with classical phase space variables $x^{[i']}_\alpha$ satisfying canonical Poisson bracket relation, $\{x_p^{[i']},x_q^{[j']}\}=if_{pqr}\delta_{[i'][j']}x_r^{[i']}$ defined by the structure constants, 
    \begin{equation}
        \hat{X}_p^{[i']}\rightarrow x_p^{[i']}-\frac{i}{2}x_q^{[i']}f_{pqr}\frac{\partial}{\partial x_r^{[i']}}
    \end{equation}
    
    \item[$\circ$] Represent the Hamiltonian and observables as functions of classical phase space variables. 
    
    \begin{equation}\hat{O}^{[i']}=\sum_p o_p \hat{X}^{[i']}_p\rightarrow O_{\mathbf{W}}(\{x\})=\sum_po_px_p^{[i']}
    \end{equation}
    with $o_p=\frac{1}{D}\mathrm{Tr}[\hat{O}\hat{X}_p^{[i']}]$, and 
    
    \begin{equation}
    \hat{H}\rightarrow H_\mathbf{W}=\sum_{[i'][j']}\mathbf{J}_{pq}^{[i'][j']}x^{[i']}_p x^{[j']}_q+\sum_{[j']}\mathbf{B}^{[j']}_p x^{[j']}_p
    \end{equation}
    where index $\mathbf{W}$ indicates that this is the Weyl symbol corresponding to symmetric operator ordering.
    
    \item[$\circ$] Find or approximate the Weyl-symbol of the initial state, i.e.\ its Wigner function. While it can assume negative values, we require that it is completely positive and thus can be thought of as a probability distribution. We present two possible definitions for the Wigner function below this implementation guide.
    
    \item[$\circ$] Solve the classical equations of motion for the phase space variables,
    \begin{equation}
    \frac{dx_p^{[i']}(t)}{dt}=-\{x_p^{[i']},H_\mathbf{W}\}=f_{pqr}\frac{\partial H_\mathbf{W}}{\partial x_q^{[i']}}x_r^{[i']}
    \end{equation}

    \item[$\circ$] Find expectation values of observables by averaging the corresponding classical functions over phase space points sampled from the Wigner function, $\langle \hat{O}(t)\rangle=\lim_{M \to \infty}\frac{1}{M}\sum_m^{M} O_\mathbf{W}(\{x(t)\}_m)$, where $M$ denotes the number of samples.
    
\end{itemize}

\subsubsection{Gaussian Wigner function}
Wurtz~et~al.~\cite{Wurtz2018} defined an approximate Gaussian Wigner function $\mathcal{W}(\{x\})$ describing the initial conditions for the system with the only requirement that the initial state factorizes between clusters such that $\mathcal{W}(\{x\})=\prod_{[i']}\mathcal{W}^{[i']}(\{x^{[i']}_\alpha\})$, where 
\begin{equation}
    \mathcal{W}^{[i']}({x^{[i']}})=\frac{1}{Z}\exp{\Big[(x_p-\rho_p^{[i']})\Sigma_{pq}^{[i']}(x_p-\rho_q^{[i']})\Big]}
\end{equation} 
is simply a Gaussian. To determine coefficients $\rho_p^{[i']}$ and $\Sigma_{pq}^{[i']}$ from the initial density matrix on cluster $[i']$, we demand (cluster index $[i']$ suppressed)
\begin{align}
    {\rm Tr}[\hat \rho\hat X_q] &= \int\!\prod_p \mathrm{d}x_p x_q \mathcal{W}(\{x\})\quad\\
    {\rm Tr}[\hat \rho(\hat X_q\hat X_r+\hat X_r\hat X_q )] &= 2\!\int\!\prod_p \mathrm{d}x_p x_q x_r \mathcal{W}(\{x\})
\end{align}
such that the moments match to second order.

\subsubsection{Discrete Wigner function}\label{sec:discrete-wigner-function}
While the Gaussian approximation of the Wigner function described above is quite general, it neglects the moments beyond the second order. dTWA on the other side can capture all moments of the single-spin observables for initial states that factorize between individual spins~\cite{Schachenmayer2015}. 
In the following, we briefly recapitulate the derivation dTWA's sampling to introduce the notation and then generalize the method to clusters of spins.

The key idea behind the dTWA is to replace the Gaussian approximation of the Wigner function with a discrete Wigner function defined via a discrete set of phase-point operators $\mathbf{\hat{A}}^{\bigotimes n}=\bigotimes_{i}^{N} \hat{A}^{[i]}$ where $\hat{A}^{[i]}$ are discrete phase-point operators that span the $SU(2)$ phase-space of the $i$th spin. These are usually defined as $\hat{A}^{[i]}_{p,q} = (\mathds{1}+\mathbf{r}(p,q)\cdot\mathbf{\hat{\sigma}})/2$, $\mathbf{\hat{\sigma}} = (\hat{\sigma}_x,\hat{\sigma}_z,\hat{\sigma}_z)$ are the Pauli matrices and $\mathbf{r}(p,q)$ denotes suitable combinations thereof (cf.~\cite{WOOTTERS19871,Schachenmayer2015,Pucci2016}): $\mathbf{r}(0,0)=(1,1,1)$, $\mathbf{r}(0,1)=(-1,-1,1)$, $\mathbf{r}(1,0)=(1,-1,-1)$ and $\mathbf{r}(1,1)=(-1,1,-1)$. In case the wavefunction factorizes, the Wigner function of the system is then given simply by the product of single spin Wigner function given by $w^{[i]}(p,q)=\braket*[1]{\hat{A}_{(p,q)}}/2$. Crucially, for all spin states pointing along one axis each value of $w^{[i]}(p,q)$ is positive and, since they sum to 1, one can interpret them as probability distribution to sample from. 
A schematic illustration is provided in Fig.~\ref{fig2}.

Considering a system of $n_c$ clusters of $n$ spins each, we again seek to describe the state by a discrete Wigner function. The main difference to before is that each local Hilbert space is represented by a copy of $SU(D)$, where $D=2^n$. In analogy to before, we introduce the phase-point operators $\hat{\mathbf{A}}^{\bigotimes n_c}=\bigotimes_{i'}^{n_c} \hat{A}_n^{[i']}$ with $\hat{A}_n^{[i']}=(\mathds{1}_D+\mathbf{r}_n^{[i']}\cdot\mathbf{X}_n)/D$, where $\mathbf{r}_n^{[i']}=(r_1^{[i']},\cdots,r_{D^2-1}^{[i']})_i$ represents a vector of $D^2-1=4^n-1$ real-valued parameters and $\mathbf{X}_n$ corresponds to a vector of the operators of the operator basis for a cluster of $n$ spins as used in cTWA. Note, we can construct the operator basis $\mathbf{X}_n$ for $n$ spins iteratively from an operator basis $\mathbf{X}_1$ for a single spin by taking tensor products
$\mathbf{X}_n = (\mathbf{X}_1 \otimes \mathds{1}, \mathds{1}\otimes\mathbf{X}_{n-1},  \mathbf{X}_1 \otimes\mathbf{X}_{n-1})$. One can construct $\mathbf{r}_n$ analogously:
\begin{equation}
    \mathbf{r}_n(\mathbf{p},\mathbf{q}) = \left[\mathbf{r}_1(p_1,q_1), \mathbf{r}_{n-1}(\mathbf{\tilde{p}},\mathbf{\tilde{q}}),  \mathbf{r}_1 (p_1,q_1) \otimes\mathbf{r}_{n-1}(\mathbf{\tilde{p}},\mathbf{\tilde{q}})\right]
\end{equation} with $\mathbf{p},\mathbf{q} \in \{0,1\}^n$ and $\mathbf{\tilde{p}}$ ($\mathbf{\tilde{q}}$) denoting the vector derived from $\mathbf{p}$ ($\mathbf{q}$) by dropping the first element.
Suppressing the index $n$ from now on, the Wigner function of a cluster is defined as before to be $w^{[i']}(\mathbf{p},\mathbf{q}) = \braket*[1]{\hat{A}_{(\mathbf{p},\mathbf{q})}}/D$.
If the quantum wavefunction factorizes between spins within a cluster, the Wigner function also factorizes and the result is essentially equivalent to the single spin case:
\begin{equation}
    w^{[i']}(\mathbf{p},\mathbf{q}) = \prod_{i}\braket[1]{\hat{A}_{(p_i,q_i)}}/2 = \prod_{i}w^{[i]}(p_i,q_i)
\end{equation}
The key difference is in the phase-point vectors $\mathbf{r}_n(\mathbf{p},\mathbf{q})$ connected to this Wigner function which now also encompass a much larger operator basis.
In summary, if the initial wave function factorizes between spins, one can simply sample the initial values for the single spin operators and compute the initial values for operators acting on multiple spins by appropriate products. For a more detailed description of the sampling process, see appendix \ref{app:sampling-discrete-Wigner-cluster}.

As a concrete example, consider a cluster of 2 spins in a Néel state $\ket{\uparrow\downarrow}$. To generate a sample, one draws the 4 values for $\braket*[1]{\hat{X}_1}$, $\braket*[1]{\hat{Y}_1}$, $\braket*[1]{\hat{X}_2}$, and $\braket*[1]{\hat{Y}_2}$ randomly from $\{-1,1\}$ and sets $\braket*[1]{\hat{Z}_1}=-\braket*[1]{\hat{Z}_2}=1$. Then rest of the correlators are computed from the products of these, e.g.\ $\braket*[1]{\hat{X}_1 \hat{Y}_2} = \braket*[1]{\hat{X}_1}\braket*[1]{\hat{Y}_2}$ and so on.
This means that the initial spin vectors are randomly drawn from one of the 16 spin configurations. All other states on the Bloch sphere can be sampled using the same configurations, followed by an appropriate rotation. 

\subsection{Clustering strategies}
The cTWA necessitates a choice of clustering of the spins. While in ordered systems, it makes sense to simply choose contiguous regions of desired size, in disordered systems it is not clear a priori that this is a reasonable choice. In this work, we evaluate two possible strategies for choosing the clustering:

\begin{itemize}
    \item \textit{Naive clustering method} (see Fig.~\ref{fig1}-b):
   In this approach, clusters of spins are formed by starting from one end of the chain and grouping together every $n$ consecutive spins.
   Thus, the resulting clusters are determined solely on the basis of this selection process, without taking into account any specific properties or interactions between the spins. 
   \item \textit{Renormalization Group clustering} [see Fig.~\ref{fig1}(c)]:
   The RG clustering strategy takes inspiration from the real space renormalization group (RSRG), also known as strong disorder renormalization group (SDRG), approaches to bond-disordered models which are used to construct approximate eigenstates~\cite{altmanUniversalDynamicsRenormalization2015,vasseurQuantumCriticalityHot2015,igloiStrongDisorderRG2018}. These methods identify the two spins sharing the strongest coupling in the system and treats their couplings to the environment in a perturbative manner. Since this procedure effectively decouples the pair and leaves the form of the remaining Hamiltonian invariant, one can readily repeat this procedure with the remaining spins until all spins are paired up. Instead of computing eigenstates, we simply use the resulting partition of spins into clusters of size two as input for the cTWA. In this way, the strong intra-pair interactions are treated fully quantum mechanically, while the effective interaction among pairs is treated semi-classically.
\end{itemize}
Based on the excellent results found by RSRG/SDRG, we expect the RG clustering to outperform the naive clustering method for strong disorder. However, the naive clustering scheme generalizes naturally to generate larger clusters, while it is unclear how to merge the clusters given by the RG scheme in a consistent manner.

\section{RESULTS}
In this section, we present the numerical results of our exploration of the quench dynamics of a disordered spin chain with long-range interactions. We compute the behavior of two dynamical observables initiated from an N\'eel state, namely the staggered magnetization and the R\'enyi entropy $S_2(t)$ evaluated over a two-spin subsystem, using the different methods detailed above and compare to results obtained with exact diagonalization (ED).
Our primary focus lies on evaluating the performance of the cluster truncated Wigner approximation (cTWA) relative to standard dTWA. To this end, we consider combinations of the aforementioned clustering schemes, the naive clustering and the one based on the strong disorder renormalization group, and the two approximations of the initial state, the Gaussian cTWA (gcTWA) and discrete cTWA (dcTWA).
Our analysis aims to shed light on how cTWA captures the intricate behavior of the system under bond disorder and long-range interactions, and to elucidate the extent to which this approach provides insights into the quantum dynamics of the system under consideration. All curves shown are obtained using 1000 disorder samples and 1000 Monte-Carlo trajectories unless specified otherwise. Disorder shots are identical across the methods. The code is freely available at \href{https://github.com/abraemer/DiscreteCTWAPaper}{GitHub}~\footnote{\href{https://github.com/abraemer/DiscreteCTWAPaper}{https://github.com/abraemer/DiscreteCTWAPaper}}. Statistical errors due to finite sample sizes of disorder and Monte Carlo trajectories are below the width of the lines.

\subsection{Bond disordered XX chain}

\begin{figure}
    \centering
    \includegraphics[width=1. \columnwidth]{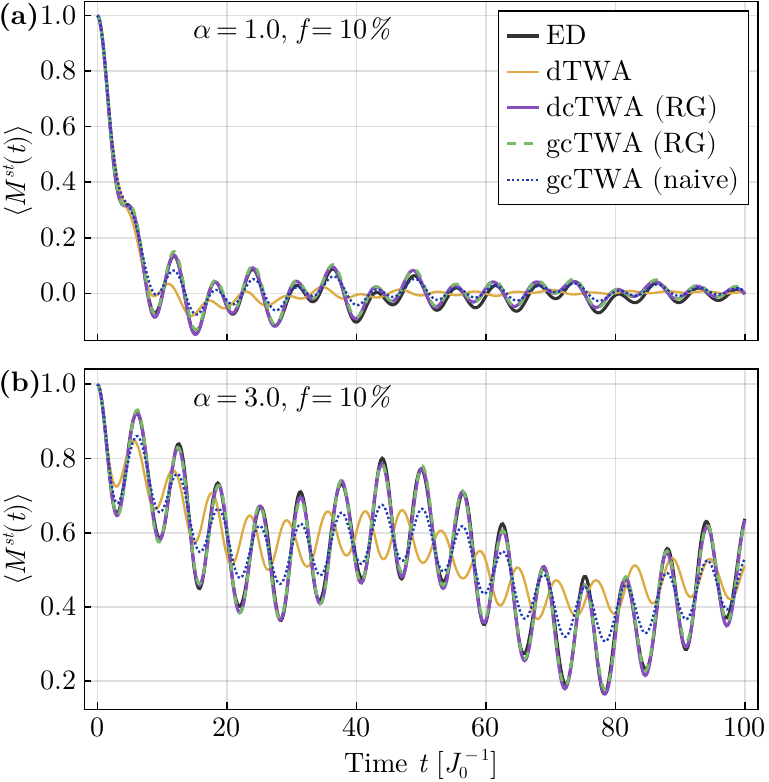}
    \caption{The disorder-averaged staggered magnetisation $\big\langle M^{\rm st}(t)\big\rangle$ is shown for XX chain of $N=16$ spins with a fixed filling of $f=10\%$. The panels show results for long-range interactions with $\alpha=1.0$ in (a) and short-range interactions with $\alpha=3.0$ in (b). The semi-classical cluster methods using the RG-inspired clustering (green, dashed and purple, solid) overlap the exact results (black, solid) almost completely. dTWA (yellow, solid) and gcTWA with naive clustering (blue, dotted) deviate already early on ($t\approx 10J_0$).}
    \label{fig3:strong-disorder-initial-comparison}
\end{figure}

\begin{figure}
    \centering
    \includegraphics[width=1. \columnwidth]{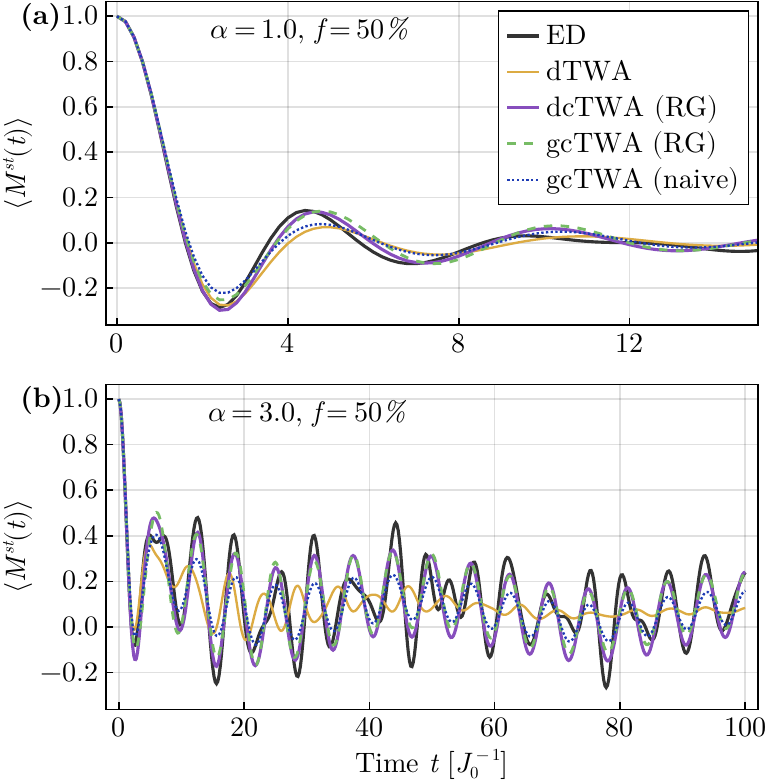}
    \caption{Same as Fig.~\ref{fig3:strong-disorder-initial-comparison}, but for filling fixed at $f=50\%$ (weak disorder regime). Here, the methods using the RG-inspired clustering show some deviation from the exact results and for $\alpha=1$ in panel (a) there are some differences between the Gaussian and discrete sampling schemes visible as well.}
    \label{fig4:weak-disorder-initial-comparison}
\end{figure}

We begin our investigation by considering a bond-disordered XX chain ($\Delta=0$). We explore various regimes by adjusting two key parameters: the power-decay exponent of the interaction, denoted as $\alpha$, and the filling fraction of the lattice, denoted as $f$, which controls the strength of the disorder. Here a low filling fraction corresponds to strong disorder, while $f=100\%$ represents a fully ordered system.

Fig.~\ref{fig3:strong-disorder-initial-comparison} shows the disorder-averaged time evolution of the staggered magnetisation $\big\langle M^{\rm st}(t) \big\rangle$ with a fixed filling of $f=10\%$, starting from the N\'eel state. The top panel corresponds to long-range interactions ($\alpha=1$), while the bottom panel corresponds to short-range interactions ($\alpha=3$). 
The staggered magnetisation starts at a value of one, which reflects the perfect order inherent in the initial N\'eel state. It then undergoes a decay, caused by the spins exchanging magnetization as system evolves. At late times, we observe equilibration to a value close zero. The general behavior is captured by all semi-classical methods.

Upon closer inspection, it becomes evident that the dTWA approach fails to accurately track the true dynamics and loses accuracy even at intermediate timescales starting around $tJ_0\approx 10$. In both cases, it predicts the location of the first oscillation approximately correctly but underestimates the amplitude. Subsequently, it systematically underestimates the amplitude of the oscillations of the staggered magnetization. Interestingly, gcTWA with the naive clustering does not fare much better. While it is generally more accurate with respect to the oscillation frequency, it also underestimates the amplitude. In contrast, both cTWA variants using the RG clustering yield essentially exact results even at late times. 
This is a very strong indicator that the dynamics is strongly shaped by the presence of strongly interacting pairs of spins where interactions among pairs are weak~\cite{Braemer2022}. With this pair model we can explain the observed curves qualitatively: It is known that dTWA is unable to correctly capture the dynamics of even a single pair and can just approximate the decay timescale (cf. App.~\ref{app:single-pair-dynamics}). If the two spins forming a strongly interacting pair are not part of the same cluster, then cTWA treats the interactions within the pair semi-classically similar to dTWA and thus faces the same problems. Consequently, using the naive clustering will result in a mixture of "correctly" and "incorrectly" chosen pairs and thus cTWA with this type of clustering provides only a slight improvement over dTWA. The RG clustering, in turn, ensures that all strongly interacting pairs are treated as clusters and thus the predictions match the exact dynamics much more closely. In turn, the high degree of agreement between cTWA with the RG clustering is also testament to the quality of pair approximation.

To further explore the efficacy of cTWA in regimes of weak disorder, we increase the filling fraction to $f=50\%$ and repeat the analysis (cf. Fig.~\ref{fig4:weak-disorder-initial-comparison}). In this regime, we do not expect the pair approximation to be accurate anymore. Indeed, for the long-range case $\alpha=1$ we find all semi-classical methods to overestimate the oscillation frequency similarly. Both RG clustering based methods predict the amplitudes almost exactly correct, while dTWA and cTWA with naive clustering again clearly underestimate it. In the more short-range case $\alpha=3$, the picture is more complex. dTWA performs worst out of all the methods and does not resolve the oscillation well and cTWA with naive clustering again essentially underestimates the amplitude. Interestingly, in this case there is a clear difference on intermediate timescales $t\approx8J_0$ between both cTWA methods with RG clustering but different choice of sampling. The discrete sampling captures the first oscillation slightly better both in position and in amplitude before at later times the prediction collapses onto the cTWA curve employing a Gaussian Wigner function. Generally, all cluster based methods still capture the dynamics qualitatively but not quite quantitatively over the whole time shown here. 

\begin{figure}
    \centering
    \includegraphics[width=1. \columnwidth]{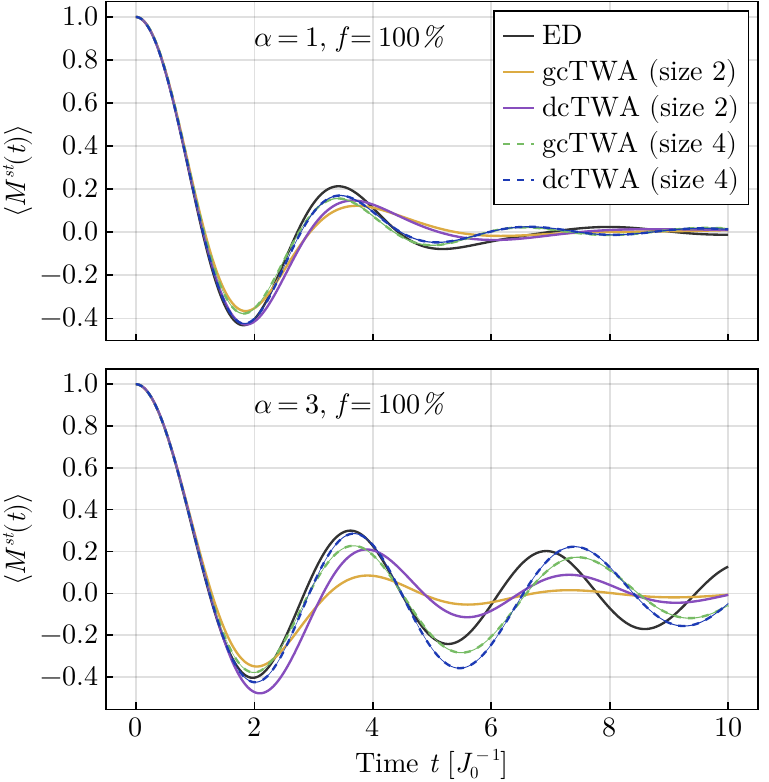}
    \caption{Comparison of sampling schemes with different cluster sizes in a clean system (filling fraction $f=100\%$). We compare TWA results for cluster sizes 2 (solid) and 4 (dashed) using the naive clustering and the different sampling schemes, Gaussian (lighter colors) and discrete (darker colors) to exact results (black, solid). Generally, cluster size 4 is more accurate than cluster size 2 and the discrete sampling scheme agrees with the exact results longer than the Gaussian sampling scheme. Other parameters are similar to Fig.\ref{fig3:strong-disorder-initial-comparison}.}
    \label{fig5:clean-disorder-initial-comparison}
\end{figure}

For a better comparison of the sampling schemes, it is instructive to examine a perfectly ordered regime by setting the filling factor to $f=100\%$. In this setting, the RG and naive clustering schemes result in the same choice of clusters and we use this opportunity to check the convergence with increasing cluster size.
Fig.~\ref{fig5:clean-disorder-initial-comparison} shows the staggered magnetisation results for systems with both long-range ($\alpha=1.0$) and short-range ($\alpha=3.0$) spin interactions and for cluster sizes $2$ and $4$. Similar to the weakly disordered case before, cluster size $2$ is insufficient to capture the relaxation dynamics quantitatively. In the short-range case ($\alpha=3$) gcTWA (cluster size 2) struggles to reproduce the oscillatory behavior, which is reflected better by dcTWA. This likely stems from the fact that this coherent dynamics comes about due to the discrete nature of the spin-$\frac{1}{2}$s which is mimicked by the discrete sampling procedure~\cite{Schachenmayer2015}. Conversely, for the long-range system ($\alpha=1$), this effect is weaker as spins hybridize more due to the stronger interactions. Interestingly, for this setting dcTWA predicts the value of the first minimum more accurately than gcTWA. Increasing the cluster size to $4$ spins improves the accuracy of both methods in both cases drastically and we don't find significant differences between the sampling schemes in the long-range case. However, for the short range case, we find the discrete sampling scheme to approximate the true amplitude of the oscillation generally better than the Gaussian scheme.

\begin{figure}
    \centering
    \includegraphics[width=1. \columnwidth]{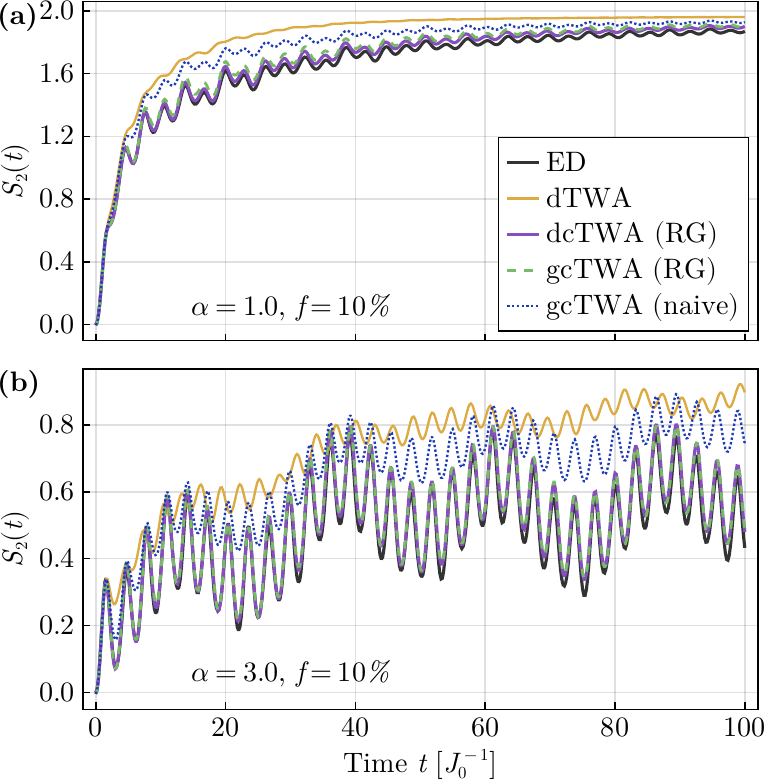}
    \caption{The average R\'enyi entropy $\big\langle S_2(t) \big\rangle$ is calculated over all possible choices of two sites. This analysis is performed with the same parameter settings as in Fig.~\ref{fig3:strong-disorder-initial-comparison}. The cTWA methods using the RG-inspired clustering (purple, solid and green, dashed) reproduce the exact entanglement dynamics (black, solid) almost exactly with only very slight deviations at late time. Whereas the gcTWA with naive clustering (blue, dotted) overestimates the entanglement and dTWA (yellow, solid) even more so.}
    \label{fig6:strong-disorder-renyi2}
\end{figure}

\begin{figure}
    \centering
    \includegraphics[width=1. \columnwidth]{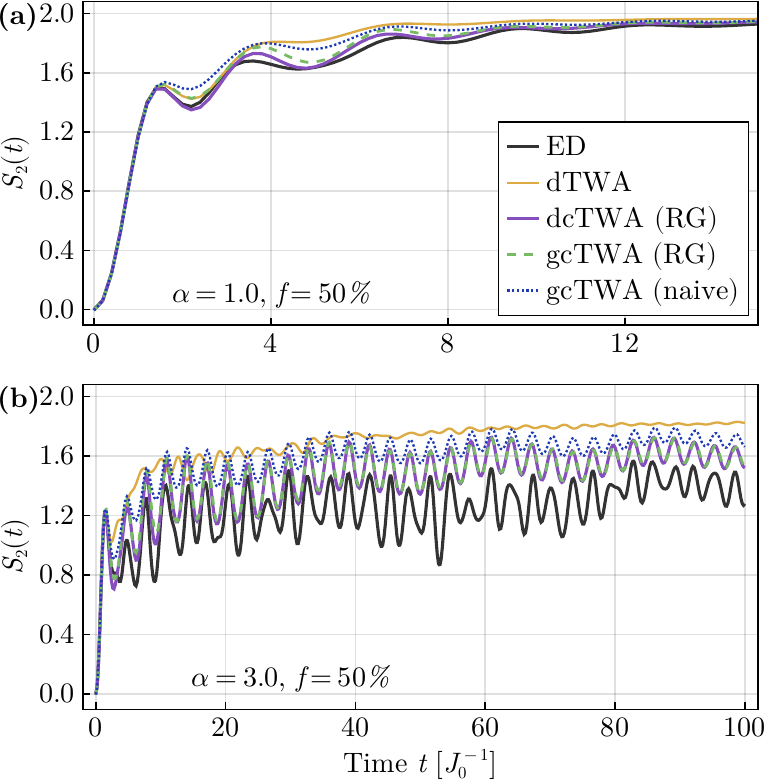}
    \caption{Same as Fig.~\ref{fig6:strong-disorder-renyi2}, but for density fixed at $f=50\%$.In comparison to Fig.~\ref{fig6:strong-disorder-renyi2}, the RG-based cTWA methods deviate from each other and also overestimate the true amount of entanglement present. In the long-range case (a) they don't capture the oscillation frequency correctly, however in the short-range case (b) they do but underestimate the amplitude.}
    \label{fig7:weak-disorder-renyi2}
\end{figure}

To extend our investigation to more complicated, non-local observables, we study the R\'enyi entropy of two-spin subsystems and assess the efficacy of the semi-classical methods under scrutiny. More specifically, we consider the average R\'enyi entropy across all possible choices of two sites.

Starting with the strongly disordered setting at $f=10\%$ in an analogy to above, Fig.~\ref{fig6:strong-disorder-renyi2} illustrates the dynamics of the average R\'enyi entropy with time in a strongly disordered setting. Since the initial state is a product state, entanglement starts at 0 for $t=0$ and then starts to increase.
We find, that generally the semi-classical methods are able to capture the dynamics across the different settings probed qualitatively, as shown in Fig.~\ref{fig6:strong-disorder-renyi2} and Fig.~\ref{fig7:weak-disorder-renyi2}. Perhaps surprisingly at first, these methods systematically overestimate the amount of entanglement present. This conundrum can be resolved, if one considers that the R\'enyi entropy is computed by estimating the expectation values of all intra-pair correlators and less correlations means more entanglement of the pair with its environment (cf.~Eq.~\ref{eq:TWA-renyi2}). The semi-classical methods miss out on some of the quantum correlations, thus tend to underestimate the total amount of correlations and consequently predict too much entanglement.
Again, the quality of the result depends significantly on the scheme. The deviations are most pronounced for dTWA and the cTWA with naively chosen clusters. Conversely, both dcTWA and gcTWA with the RG clustering scheme approximate the exact dynamics very closely and only overestimate the entanglement by a few percent at late times.

In summary, we find that cTWA may offer tremendous improvements over the simpler dTWA. However, the improvement depends strongly on the choice of clusters. If the clustering does not respect the underlying physics, as is the case for the naive clustering strategy, cTWA showed only a very minor increase in accuracy. On the other hand, if the dominant physical processes are mostly contained within the chosen clusters, as is the case with the RG inspired clustering, cTWA can describe the dynamics of the system over all time intervals almost exactly. The results obtained with the Gaussian Wigner function were very similar compared to the discrete sampling with a slight advantage in favor of the discrete scheme for ordered, short-range systems.

\subsection{Bond disordered XXZ chain}
\begin{figure}
    \centering
    \includegraphics[width=1.\columnwidth]{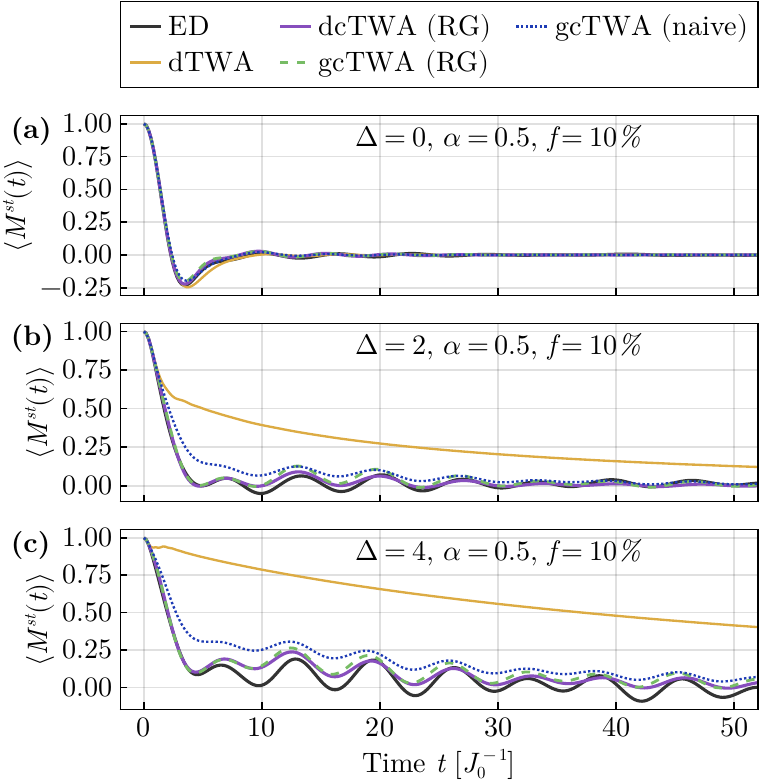}
    \caption{The disorder-averaged staggered magnetisation $\big\langle M^{\rm st}(t) \big\rangle$ is shown for XXZ chain of size $N=16$ with  $\alpha=0.5$ at $f=10\%$. Different panels are shown different $\Delta$. Results of cTWA, dTWA and ED are shown with solid-blue, dotted-green, and solid-blue, respectively.}
    \label{fig8}
\end{figure}

\begin{figure}
    \centering
    \includegraphics[width=1.\columnwidth]{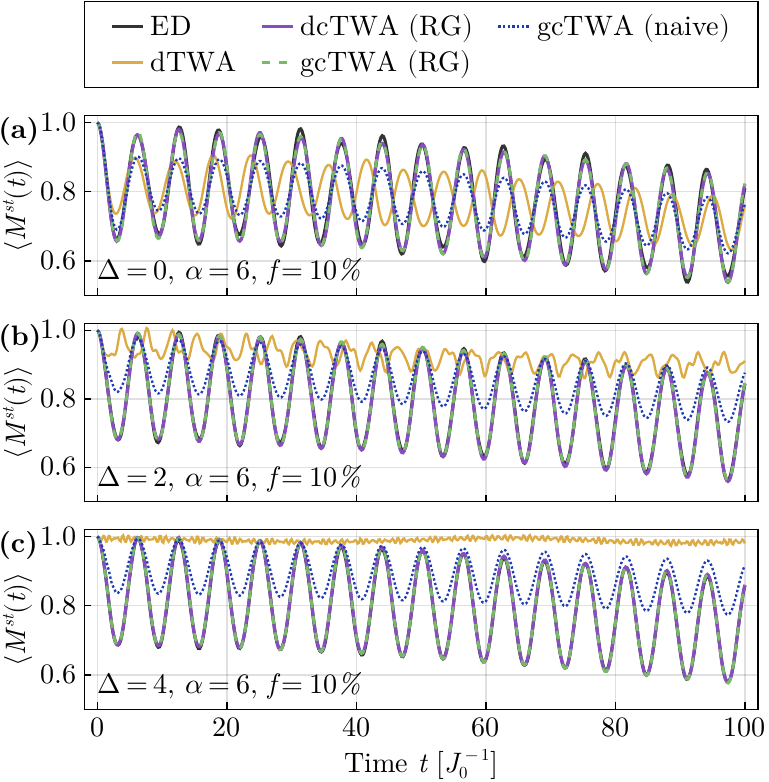}
    \caption{Same as Fig.\ref{fig8} but for $\alpha=6.0$}
    \label{fig9}
\end{figure}

\begin{figure}
    \centering
    \includegraphics[width=1.\columnwidth]{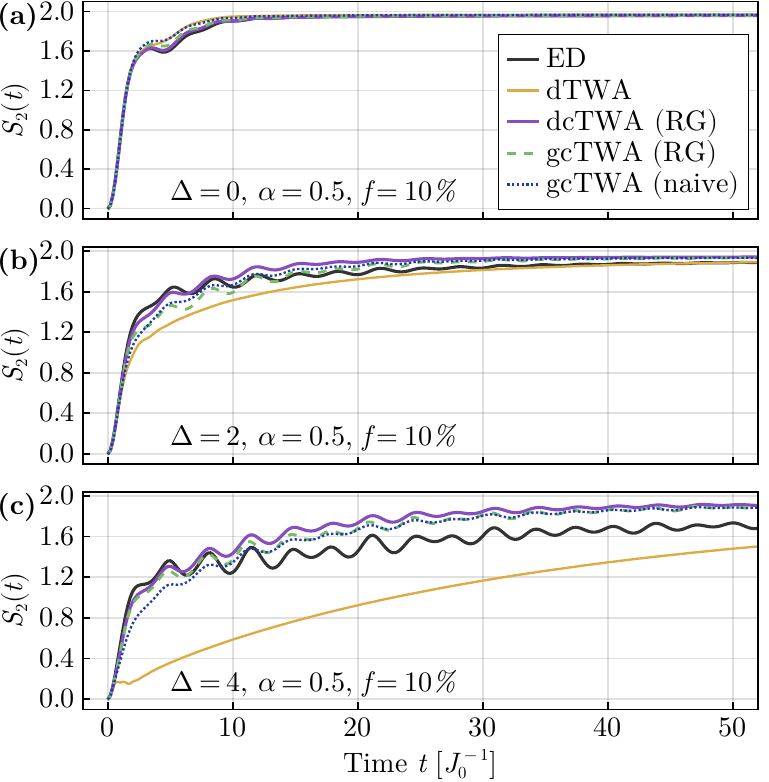}
    \caption{The average R\'enyi entropy $\big\langle S_2(t) \big\rangle$ is calculated over all possible pairs of two sites. This analysis is performed with the same parameter settings as in Fig.~\ref{fig8}.}
    \label{fig10:renyi-ising-term-long-range}
\end{figure}

\begin{figure}
    \centering
    \includegraphics[width=1.\columnwidth]{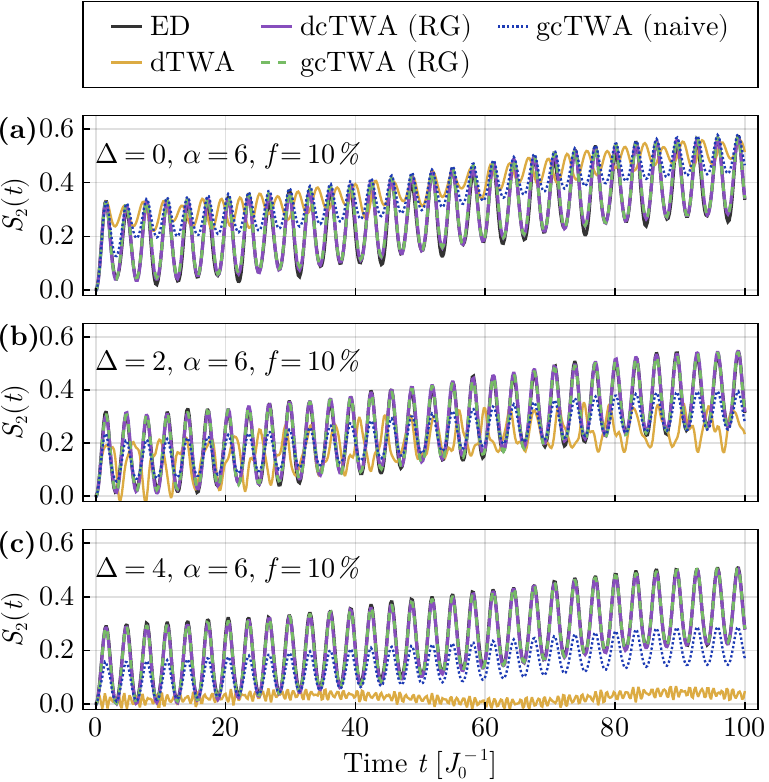}
    \caption{Same as Fig.\ref{fig10:renyi-ising-term-long-range} but for $\alpha=6.0$}
    \label{fig11:renyi-ising-term-short-range}
\end{figure}

In this section, we investigate the role of the anisotropy parameter $\Delta$ in the dynamics of the system. In the pair picture, a strong anisotropy increases the energy gap between the sectors of different absolute z-magnetization. This does not alter the dynamics of a single pair initialized in a Neel state, because dynamics is fully contained within the sector of zero magnetization. As such, we expect the choice of clusters to have a large impact on the quality of the approximation whenever the dynamics is heavily dominated by pair dynamics. To expand the domain of our study, we employ $\alpha=0.5$ to evaluate settings with even more long-range interactions, which in principle should play to TWA's strengths. In the short range case, we chose $\alpha=6$, as motivated by the typical interaction exponent of Van-der-Waals interactions in Rydberg atoms which are a possible platform to implement XXZ Heisenberg models (see e.g.~\cite{geierFloquetHamiltonianEngineering2021}). We note however, that the qualitative differences to $\alpha=3$ are minor.

Figures \ref{fig8} and \ref{fig9} show the dynamics of the staggered magnetisation under long-range ($\alpha=0.5$) and short-range ($\alpha=6$) interactions, respectively. Starting again from the Néel state, we examine the evolution of the staggered magnetization by varying $\Delta$, assessing how these adjustments affect the dynamics and how well semi-classical methods approximate the true dynamics.
At $\Delta=0$, all semi-classical methods give results matching the exact solution over almost the entire time scale for the long-range system ($\alpha=0.5$), while in the short-range system ($\alpha=6$) only the cTWA simulations using the RG-inspired clustering provide accurate results. dTWA performs worst by predicting oscillation with both wrong amplitude and frequency. gcTWA with naive clustering improves upon this due to the inclusion of more quantum correlations which results in a correct prediction of the frequency. 

Increasing the Ising interaction [cf. Figures~\ref{fig8} and \ref{fig9} (b) and (c)] does not alter the exact dynamics qualitatively, but dTWA increasingly deviates from the exact results vastly underestimating the rate of the initial decay. For the short range system and $\Delta=4$ the decay is almost completely suppressed. By contrast, gcTWA with naive clustering yields significantly better results than dTWA. For both systems, the gcTWA prediction qualitatively matches the exact data but is offset by an increasing amount with increasing $\Delta$.
Interestingly, both cTWA variants using the RG-inspired clustering, match the reference rather closely over the entire time domain except for intermediate times for $\Delta=4$ in the long-range system $\alpha=0.5$, where the fluctuations are not reproduced exactly. Surprisingly, this hints at pairs still playing an important role for the dynamics in spite of the quite long-range interactions. For the short-range interactions with strong disorder, the precise match is no surprise as the dynamics is governed by pairs of spins on adjacent lattice sites in this regime.

Again, we we employ the semi-classical methods to also extract the average pair R\'enyi entropy and compare to exact results. Starting with the long-range scenario, $\alpha=0.5$, we find for $\Delta=0$, all semi-classical approaches to converge to the true dynamics approximately (Fig. \ref{fig10:renyi-ising-term-long-range}). Increasing $\Delta$, we can see again how dTWA fails to capture the essential processes and predict much too slow dynamics (roughly one order of magnitude too slow). gcTWA with naive clustering fares rather well and only slightly underestimates the initial rise for $\Delta=2$ very similar to gcTWA with RG-inspired clustering. Most interesting are the differences between gcTWA and dcTWA (both with RG clustering) since in this setting both methods seem to converge to slightly different results with dcTWA following the exact curve more closely at intermediate times (up to $t\approx10J_0$). At late times all cTWA methods overestimate the amount of entanglement present. This trend continues for $\Delta=4$ where the discrepancy is enhanced for all methods.

For the short-range interacting systems, we find that cTWA schemes based on the RG clustering to be in excellent agreement with the exact results, while the naive clustering gcTWA and dTWA fail to capture the dynamics. This stark contrast to the long-range interacting systems likely originates in the much broader distribution of couplings caused by the much shorter interaction range. Since the RG clustering scheme incorporates the strongest of the relevant couplings, the system can show deviations only at very late timescales.

In order to understand, why dTWA struggles to accurately capture the dynamics of even a single pair of spin interacting via an XXZ Hamiltonian $H = J (\sigma^1_x\sigma^2_x + \sigma^1_y\sigma^2_y) + \Delta\sigma^1_z\sigma^2_z$ we need to consider its spectrum. Eigenstates of $H$ are the maximally entangled Bell states $\ket{\pm}=(\ket{\uparrow\downarrow}\pm\ket{\downarrow\uparrow})/\sqrt{2}$ at energies $E_\pm = \pm 2J - \Delta$ and the polarized states $\ket{\uparrow\uparrow}$ and $\ket{\downarrow\downarrow}$ with energy $E_p = \Delta$. Taking the Neel state $\ket{\uparrow\downarrow} = (\ket{+}+\ket{-})/\sqrt{2}$ as initial state, the exact quantum dynamics only populates the two maximally entangled eigenstates. Since their energetic splitting depends on $J$ only, the exact dynamics is independent of $\Delta$ and just encompasses the the coherent flipping of both spins $\ket{\uparrow\downarrow}\leftrightarrow\ket{\downarrow\uparrow}$. dTWA essentially has access only to single body-terms and thus needs to approximate this process by two steps which will couple to the polarized states. This gives an intuitive understanding of the dependence on $\Delta$ for dTWA. The precise nature of this relation is quite intricate and not akin to, e.g., a two-photon transition. For further analysis of the 2 spin case with dTWA, we refer to Appendix~\ref{app:single-pair-dynamics}.
At this point, we want to remark that even cTWA of course captures the dynamics exactly if the two spins are part of the same cluster, but still the state is not represented exactly at all times. As we have shown in Appendix~\ref{app:sampling-discrete-Wigner-cluster}, dcTWA can only represent states where the Wigner function is non-negative but one can ensure that all observables within a cluster have correct means and only higher moments deviate.

In summary, we find that even in very long-range systems and for strong Ising interactions the RG-inspired clustering yields quite accurate results at early and intermediate times. At late times, we see some deviations that increase with the strength of the Ising couplings which likely signals the breakdown of the pair approximation in this regime. For short-range interactions, the cTWA methods with RG clustering yield basically exact results in all cases studied here. Conversely, dTWA and gcTWA with the naive clustering strategy struggle due to the competition of the Ising and hopping interactions. We did not see a significant difference between discrete and Gaussian sampling in these settings.

\subsection{Statistical Error Analysis}
\begin{figure}
    \centering
    \includegraphics[width=1.\columnwidth]{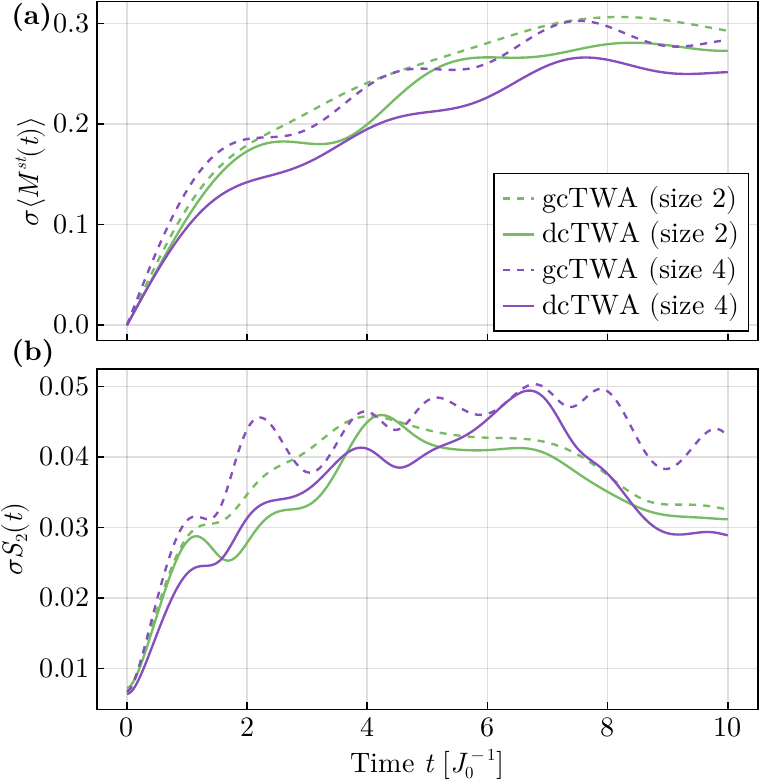}
    \caption{Error analysis for a single shot with the parameters $\alpha=1$, $f=10\%$, $\Delta=0$ and $10,000$ trajectories. We compute the standard deviation of (a) the staggered magnetization and (b) average pair Renyi entropy estimated from batches of $100$ trajectories. We use naive clustering to compare both cluster sizes 2 (green lines) and 4 (purple lines) as well as Gaussian sampling (dashed) and discrete sampling (solid).}
    \label{fig12:stddev-sampling-schemes}
\end{figure}

To highlight the merits of the discrete sampling scheme, we study the convergence of the staggered magnetization Monte-Carlo samples by extracting standard deviation of the staggered magnetization across 10.000 trajectories of a single disorder shot. While previous analyses did not show large differences in result between the sampling schemes, Fig.~\ref{fig12:stddev-sampling-schemes}(a) reveals the higher accuracy of the discrete sampling schemes which leads to a reduced number of samples required to achieve a given level of precision. Averaged over the timescale shown, we report approximately $8\%$ smaller standard deviation for dcTWA with cluster size 2 and $15\%$ reduction for cluster size 4. This translate to approximately $16\%$ respectively $28\%$ fewer trajectories needed to achieve similar levels of accuracy.

We repeat this analysis for the Renyi entropy, where we estimate the standard deviation from 100 sets of 100 trajectories each [cf. Fig.~\ref{fig12:stddev-sampling-schemes}(b)]. Again by averaging, we find a similar reduction of $14\%$ and $29\%$ reduction in standard deviation for cluster sizes 2 and 4 respectively.

\section{Conclusion}
In this study, we conducted a comprehensive numerical investigation focusing on the Cluster Truncated Wigner Approximation (cTWA) for modelling quench dynamics in disordered spin chains with power-law interactions. Through comparisons with the Discrete Truncated Wigner Approximation (dTWA) and the Exact Diagonalisation (ED), we explored the performance of the cTWA on different time scales relevant to quench experiments and studied the influence of the choice of clusters on the prediction. Additionally, we introduced a new sampling scheme  for generating Monte-Carlo trajectories which extends the discrete Wigner function known from dTWA to the realm of cTWA. Our analysis included both the XX and XXZ models with bond disorder initiated from a N\'eel state, and calculated dynamical observables such as staggered magnetization and two-site R\'enyi entropy.

We found that while cTWA generally yields improved accuracy compared to dTWA, the choice of clusters strongly impacts the results. Our results in the context of bond disorder show that a clustering strategy inspired by the strong disorder renormalisation group could yield astonishingly precise results in the presence of strong disorder, while still being very accurate even for quite long-range interactions, weak disorder and long times. XXZ models featuring strong Ising interactions were found to be challenging for all semi-classical methods presented here and we conclude that likely larger clusters are needed to capture the relevant physical processes accurately. In all of these systems, we only found minor differences between the Gaussian and discrete sampling schemes in situations were the results were not converged in cluster size. However, a closer study of the statistical properties revealed the discrete sampling to exhibit smaller intrinsic Monte-Carlo shot noise.

In conclusion, our study provides valuable insights into the effectiveness of cTWA in studying quench dynamics in bond-disordered spin systems. If using the correct clustering strategy, even clusters of size 2 yield close to exact results, where single spin dTWA fails. Additionally, we conclude that the discrete sampling strategy introduced here is generally preferrable to the Gaussian approximation due to less Monte-Carlo shot noise and somewhat simpler implementation. Overall, our results highlight the potential of cTWA and its variants, such as dcTWA, as powerful tools for studying the complex dynamics of bond-disordered quantum systems.

\section{Acknowledgment}
We thank O. Egorov for providing computational resources.
We acknowledge State of Thuringia for funding of projects "2019 FGI 0017" (PhyClus) and "2022 FGI 0004" (Q-PHOC3).
A.B. acknowledges support by the Deutsche Forschungsgemeinschaft 
(DFG, German Research Foundation) within the Collaborative Research Center SFB1225 (ISOQUANT). 
For the numerical work, we used the Julia programming language \cite{bezansonJuliaFreshApproach2017} and the following packages/tools: \texttt{DrWatson.jl}~\cite{datserisDrWatsonPerfectSidekick2020}, \texttt{Pluto.jl}~\cite{fonsvanderplasFonspPlutoJl2024}, \texttt{DifferentialEquations.jl}~\cite{rackauckas2017differentialequations} with Verner's integration schemes~\cite{vernerNumericallyOptimalRunge2010}, \texttt{Makie.jl}~\cite{danischMakieJlFlexible2021}, \texttt{DataFrames.jl}~\cite{bouchet-valatDataframesJlFlexible2023}.

\bibliography{main}

\begin{thebibliography}{65}%
\makeatletter
\providecommand \@ifxundefined [1]{%
 \@ifx{#1\undefined}
}%
\providecommand \@ifnum [1]{%
 \ifnum #1\expandafter \@firstoftwo
 \else \expandafter \@secondoftwo
 \fi
}%
\providecommand \@ifx [1]{%
 \ifx #1\expandafter \@firstoftwo
 \else \expandafter \@secondoftwo
 \fi
}%
\providecommand \natexlab [1]{#1}%
\providecommand \enquote  [1]{``#1''}%
\providecommand \bibnamefont  [1]{#1}%
\providecommand \bibfnamefont [1]{#1}%
\providecommand \citenamefont [1]{#1}%
\providecommand \href@noop [0]{\@secondoftwo}%
\providecommand \href [0]{\begingroup \@sanitize@url \@href}%
\providecommand \@href[1]{\@@startlink{#1}\@@href}%
\providecommand \@@href[1]{\endgroup#1\@@endlink}%
\providecommand \@sanitize@url [0]{\catcode `\\12\catcode `\$12\catcode `\&12\catcode `\#12\catcode `\^12\catcode `\_12\catcode `\%12\relax}%
\providecommand \@@startlink[1]{}%
\providecommand \@@endlink[0]{}%
\providecommand \url  [0]{\begingroup\@sanitize@url \@url }%
\providecommand \@url [1]{\endgroup\@href {#1}{\urlprefix }}%
\providecommand \urlprefix  [0]{URL }%
\providecommand \Eprint [0]{\href }%
\providecommand \doibase [0]{https://doi.org/}%
\providecommand \selectlanguage [0]{\@gobble}%
\providecommand \bibinfo  [0]{\@secondoftwo}%
\providecommand \bibfield  [0]{\@secondoftwo}%
\providecommand \translation [1]{[#1]}%
\providecommand \BibitemOpen [0]{}%
\providecommand \bibitemStop [0]{}%
\providecommand \bibitemNoStop [0]{.\EOS\space}%
\providecommand \EOS [0]{\spacefactor3000\relax}%
\providecommand \BibitemShut  [1]{\csname bibitem#1\endcsname}%
\let\auto@bib@innerbib\@empty
\bibitem [{\citenamefont {Anderson}(1958)}]{Anderson1958}%
  \BibitemOpen
  \bibfield  {author} {\bibinfo {author} {\bibfnamefont {P.~W.}\ \bibnamefont {Anderson}},\ }\bibfield  {title} {\bibinfo {title} {Absence of diffusion in certain random lattices},\ }\href {https://doi.org/10.1103/PhysRev.109.1492} {\bibfield  {journal} {\bibinfo  {journal} {Phys. Rev.}\ }\textbf {\bibinfo {volume} {109}},\ \bibinfo {pages} {1492} (\bibinfo {year} {1958})}\BibitemShut {NoStop}%
\bibitem [{\citenamefont {Mott}(1976)}]{MOTT1976}%
  \BibitemOpen
  \bibfield  {author} {\bibinfo {author} {\bibfnamefont {N.~F.}\ \bibnamefont {Mott}},\ }\bibfield  {title} {\bibinfo {title} {Impurity band conduction. experiment and theory the metal-insulator transition in an impurity band.},\ }\href {https://doi.org/10.1051/jphyscol:1976453} {\bibfield  {journal} {\bibinfo  {journal} {Le Journal de Physique Colloques}\ }\textbf {\bibinfo {volume} {37}},\ \bibinfo {pages} {C4} (\bibinfo {year} {1976})}\BibitemShut {NoStop}%
\bibitem [{\citenamefont {Kettemann}(2023)}]{KETTEMANN2023}%
  \BibitemOpen
  \bibfield  {author} {\bibinfo {author} {\bibfnamefont {S.}~\bibnamefont {Kettemann}},\ }\bibfield  {title} {\bibinfo {title} {Towards a comprehensive theory of metal–insulator transitions in doped semiconductors},\ }\href {https://doi.org/https://doi.org/10.1016/j.aop.2023.169306} {\bibfield  {journal} {\bibinfo  {journal} {Annals of Physics}\ }\textbf {\bibinfo {volume} {456}},\ \bibinfo {pages} {169306} (\bibinfo {year} {2023})}\BibitemShut {NoStop}%
\bibitem [{\citenamefont {Cesare}\ \emph {et~al.}(1990)\citenamefont {Cesare}, \citenamefont {Lukierska-Walasek}, \citenamefont {Rabuffo},\ and\ \citenamefont {Walasek}}]{Cesare1990}%
  \BibitemOpen
  \bibfield  {author} {\bibinfo {author} {\bibfnamefont {L.~D.}\ \bibnamefont {Cesare}}, \bibinfo {author} {\bibfnamefont {K.}~\bibnamefont {Lukierska-Walasek}}, \bibinfo {author} {\bibfnamefont {I.}~\bibnamefont {Rabuffo}},\ and\ \bibinfo {author} {\bibfnamefont {K.}~\bibnamefont {Walasek}},\ }\bibfield  {title} {\bibinfo {title} {Two-spin cluster approach to the infinite-range quantum transverse {XY} spin-glass model},\ }\href {https://doi.org/10.1016/0375-9601(90)90368-x} {\bibfield  {journal} {\bibinfo  {journal} {Physics Letters A}\ }\textbf {\bibinfo {volume} {145}},\ \bibinfo {pages} {291} (\bibinfo {year} {1990})}\BibitemShut {NoStop}%
\bibitem [{\citenamefont {De~Cesare}\ \emph {et~al.}(1992)\citenamefont {De~Cesare}, \citenamefont {Lukierska-Walasek}, \citenamefont {Rabuffo},\ and\ \citenamefont {Walasek}}]{Cesare1992}%
  \BibitemOpen
  \bibfield  {author} {\bibinfo {author} {\bibfnamefont {L.}~\bibnamefont {De~Cesare}}, \bibinfo {author} {\bibfnamefont {K.}~\bibnamefont {Lukierska-Walasek}}, \bibinfo {author} {\bibfnamefont {I.}~\bibnamefont {Rabuffo}},\ and\ \bibinfo {author} {\bibfnamefont {K.}~\bibnamefont {Walasek}},\ }\bibfield  {title} {\bibinfo {title} {Cavity-fields approach to quantum xy spin-glass models in a transverse field},\ }\href {https://doi.org/10.1103/PhysRevB.45.1041} {\bibfield  {journal} {\bibinfo  {journal} {Phys. Rev. B}\ }\textbf {\bibinfo {volume} {45}},\ \bibinfo {pages} {1041} (\bibinfo {year} {1992})}\BibitemShut {NoStop}%
\bibitem [{\citenamefont {de~Paz}\ \emph {et~al.}(2013)\citenamefont {de~Paz}, \citenamefont {Sharma}, \citenamefont {Chotia}, \citenamefont {Mar\'echal}, \citenamefont {Huckans}, \citenamefont {Pedri}, \citenamefont {Santos}, \citenamefont {Gorceix}, \citenamefont {Vernac},\ and\ \citenamefont {Laburthe-Tolra}}]{Sharma2013}%
  \BibitemOpen
  \bibfield  {author} {\bibinfo {author} {\bibfnamefont {A.}~\bibnamefont {de~Paz}}, \bibinfo {author} {\bibfnamefont {A.}~\bibnamefont {Sharma}}, \bibinfo {author} {\bibfnamefont {A.}~\bibnamefont {Chotia}}, \bibinfo {author} {\bibfnamefont {E.}~\bibnamefont {Mar\'echal}}, \bibinfo {author} {\bibfnamefont {J.~H.}\ \bibnamefont {Huckans}}, \bibinfo {author} {\bibfnamefont {P.}~\bibnamefont {Pedri}}, \bibinfo {author} {\bibfnamefont {L.}~\bibnamefont {Santos}}, \bibinfo {author} {\bibfnamefont {O.}~\bibnamefont {Gorceix}}, \bibinfo {author} {\bibfnamefont {L.}~\bibnamefont {Vernac}},\ and\ \bibinfo {author} {\bibfnamefont {B.}~\bibnamefont {Laburthe-Tolra}},\ }\bibfield  {title} {\bibinfo {title} {Nonequilibrium quantum magnetism in a dipolar lattice gas},\ }\href {https://doi.org/10.1103/PhysRevLett.111.185305} {\bibfield  {journal} {\bibinfo  {journal} {Phys. Rev. Lett.}\ }\textbf {\bibinfo {volume} {111}},\ \bibinfo {pages} {185305} (\bibinfo {year} {2013})}\BibitemShut {NoStop}%
\bibitem [{\citenamefont {Yan}\ \emph {et~al.}(2013)\citenamefont {Yan}, \citenamefont {Moses}, \citenamefont {Gadway}, \citenamefont {Covey}, \citenamefont {Hazzard}, \citenamefont {Rey}, \citenamefont {Jin},\ and\ \citenamefont {Ye}}]{Yan2013}%
  \BibitemOpen
  \bibfield  {author} {\bibinfo {author} {\bibfnamefont {B.}~\bibnamefont {Yan}}, \bibinfo {author} {\bibfnamefont {S.~A.}\ \bibnamefont {Moses}}, \bibinfo {author} {\bibfnamefont {B.}~\bibnamefont {Gadway}}, \bibinfo {author} {\bibfnamefont {J.~P.}\ \bibnamefont {Covey}}, \bibinfo {author} {\bibfnamefont {K.~R.~A.}\ \bibnamefont {Hazzard}}, \bibinfo {author} {\bibfnamefont {A.~M.}\ \bibnamefont {Rey}}, \bibinfo {author} {\bibfnamefont {D.~S.}\ \bibnamefont {Jin}},\ and\ \bibinfo {author} {\bibfnamefont {J.}~\bibnamefont {Ye}},\ }\bibfield  {title} {\bibinfo {title} {Observation of dipolar spin-exchange interactions with lattice-confined polar molecules},\ }\href {https://doi.org/10.1038/nature12483} {\bibfield  {journal} {\bibinfo  {journal} {Nature}\ }\textbf {\bibinfo {volume} {501}},\ \bibinfo {pages} {521} (\bibinfo {year} {2013})}\BibitemShut {NoStop}%
\bibitem [{\citenamefont {Fersterer}\ \emph {et~al.}(2019)\citenamefont {Fersterer}, \citenamefont {Safavi-Naini}, \citenamefont {Zhu}, \citenamefont {Gabardos}, \citenamefont {Lepoutre}, \citenamefont {Vernac}, \citenamefont {Laburthe-Tolra}, \citenamefont {Blakie},\ and\ \citenamefont {Rey}}]{Fersterer2019}%
  \BibitemOpen
  \bibfield  {author} {\bibinfo {author} {\bibfnamefont {P.}~\bibnamefont {Fersterer}}, \bibinfo {author} {\bibfnamefont {A.}~\bibnamefont {Safavi-Naini}}, \bibinfo {author} {\bibfnamefont {B.}~\bibnamefont {Zhu}}, \bibinfo {author} {\bibfnamefont {L.}~\bibnamefont {Gabardos}}, \bibinfo {author} {\bibfnamefont {S.}~\bibnamefont {Lepoutre}}, \bibinfo {author} {\bibfnamefont {L.}~\bibnamefont {Vernac}}, \bibinfo {author} {\bibfnamefont {B.}~\bibnamefont {Laburthe-Tolra}}, \bibinfo {author} {\bibfnamefont {P.~B.}\ \bibnamefont {Blakie}},\ and\ \bibinfo {author} {\bibfnamefont {A.~M.}\ \bibnamefont {Rey}},\ }\bibfield  {title} {\bibinfo {title} {Dynamics of an itinerant spin-3 atomic dipolar gas in an optical lattice},\ }\href {https://doi.org/10.1103/PhysRevA.100.033609} {\bibfield  {journal} {\bibinfo  {journal} {Phys. Rev. A}\ }\textbf {\bibinfo {volume} {100}},\ \bibinfo {pages} {033609} (\bibinfo {year} {2019})}\BibitemShut {NoStop}%
\bibitem [{\citenamefont {Zeiher}\ \emph {et~al.}(2017)\citenamefont {Zeiher}, \citenamefont {Choi}, \citenamefont {Rubio-Abadal}, \citenamefont {Pohl}, \citenamefont {van Bijnen}, \citenamefont {Bloch},\ and\ \citenamefont {Gross}}]{Zeiher2017}%
  \BibitemOpen
  \bibfield  {author} {\bibinfo {author} {\bibfnamefont {J.}~\bibnamefont {Zeiher}}, \bibinfo {author} {\bibfnamefont {J.-y.}\ \bibnamefont {Choi}}, \bibinfo {author} {\bibfnamefont {A.}~\bibnamefont {Rubio-Abadal}}, \bibinfo {author} {\bibfnamefont {T.}~\bibnamefont {Pohl}}, \bibinfo {author} {\bibfnamefont {R.}~\bibnamefont {van Bijnen}}, \bibinfo {author} {\bibfnamefont {I.}~\bibnamefont {Bloch}},\ and\ \bibinfo {author} {\bibfnamefont {C.}~\bibnamefont {Gross}},\ }\bibfield  {title} {\bibinfo {title} {Coherent many-body spin dynamics in a long-range interacting ising chain},\ }\href {https://doi.org/10.1103/PhysRevX.7.041063} {\bibfield  {journal} {\bibinfo  {journal} {Phys. Rev. X}\ }\textbf {\bibinfo {volume} {7}},\ \bibinfo {pages} {041063} (\bibinfo {year} {2017})}\BibitemShut {NoStop}%
\bibitem [{\citenamefont {Bernien}\ \emph {et~al.}(2017)\citenamefont {Bernien}, \citenamefont {Schwartz}, \citenamefont {Keesling}, \citenamefont {Levine}, \citenamefont {Omran}, \citenamefont {Pichler}, \citenamefont {Choi}, \citenamefont {Zibrov}, \citenamefont {Endres}, \citenamefont {Greiner}, \citenamefont {Vuleti{\'{c}}},\ and\ \citenamefont {Lukin}}]{Bernien2017}%
  \BibitemOpen
  \bibfield  {author} {\bibinfo {author} {\bibfnamefont {H.}~\bibnamefont {Bernien}}, \bibinfo {author} {\bibfnamefont {S.}~\bibnamefont {Schwartz}}, \bibinfo {author} {\bibfnamefont {A.}~\bibnamefont {Keesling}}, \bibinfo {author} {\bibfnamefont {H.}~\bibnamefont {Levine}}, \bibinfo {author} {\bibfnamefont {A.}~\bibnamefont {Omran}}, \bibinfo {author} {\bibfnamefont {H.}~\bibnamefont {Pichler}}, \bibinfo {author} {\bibfnamefont {S.}~\bibnamefont {Choi}}, \bibinfo {author} {\bibfnamefont {A.~S.}\ \bibnamefont {Zibrov}}, \bibinfo {author} {\bibfnamefont {M.}~\bibnamefont {Endres}}, \bibinfo {author} {\bibfnamefont {M.}~\bibnamefont {Greiner}}, \bibinfo {author} {\bibfnamefont {V.}~\bibnamefont {Vuleti{\'{c}}}},\ and\ \bibinfo {author} {\bibfnamefont {M.~D.}\ \bibnamefont {Lukin}},\ }\bibfield  {title} {\bibinfo {title} {Probing many-body dynamics on a 51-atom quantum simulator},\ }\href {https://doi.org/10.1038/nature24622} {\bibfield  {journal} {\bibinfo  {journal} {Nature}\ }\textbf {\bibinfo {volume}
  {551}},\ \bibinfo {pages} {579} (\bibinfo {year} {2017})}\BibitemShut {NoStop}%
\bibitem [{\citenamefont {Keesling}\ \emph {et~al.}(2019)\citenamefont {Keesling}, \citenamefont {Omran}, \citenamefont {Levine}, \citenamefont {Bernien}, \citenamefont {Pichler}, \citenamefont {Choi}, \citenamefont {Samajdar}, \citenamefont {Schwartz}, \citenamefont {Silvi}, \citenamefont {Sachdev}, \citenamefont {Zoller}, \citenamefont {Endres}, \citenamefont {Greiner}, \citenamefont {Vuleti{\'{c}}},\ and\ \citenamefont {Lukin}}]{Keesling2019}%
  \BibitemOpen
  \bibfield  {author} {\bibinfo {author} {\bibfnamefont {A.}~\bibnamefont {Keesling}}, \bibinfo {author} {\bibfnamefont {A.}~\bibnamefont {Omran}}, \bibinfo {author} {\bibfnamefont {H.}~\bibnamefont {Levine}}, \bibinfo {author} {\bibfnamefont {H.}~\bibnamefont {Bernien}}, \bibinfo {author} {\bibfnamefont {H.}~\bibnamefont {Pichler}}, \bibinfo {author} {\bibfnamefont {S.}~\bibnamefont {Choi}}, \bibinfo {author} {\bibfnamefont {R.}~\bibnamefont {Samajdar}}, \bibinfo {author} {\bibfnamefont {S.}~\bibnamefont {Schwartz}}, \bibinfo {author} {\bibfnamefont {P.}~\bibnamefont {Silvi}}, \bibinfo {author} {\bibfnamefont {S.}~\bibnamefont {Sachdev}}, \bibinfo {author} {\bibfnamefont {P.}~\bibnamefont {Zoller}}, \bibinfo {author} {\bibfnamefont {M.}~\bibnamefont {Endres}}, \bibinfo {author} {\bibfnamefont {M.}~\bibnamefont {Greiner}}, \bibinfo {author} {\bibfnamefont {V.}~\bibnamefont {Vuleti{\'{c}}}},\ and\ \bibinfo {author} {\bibfnamefont {M.~D.}\ \bibnamefont {Lukin}},\ }\bibfield  {title} {\bibinfo {title} {Quantum
  kibble{\textendash}zurek mechanism and critical dynamics on a programmable rydberg simulator},\ }\href {https://doi.org/10.1038/s41586-019-1070-1} {\bibfield  {journal} {\bibinfo  {journal} {Nature}\ }\textbf {\bibinfo {volume} {568}},\ \bibinfo {pages} {207} (\bibinfo {year} {2019})}\BibitemShut {NoStop}%
\bibitem [{\citenamefont {Islam}\ \emph {et~al.}(2013)\citenamefont {Islam}, \citenamefont {Senko}, \citenamefont {Campbell}, \citenamefont {Korenblit}, \citenamefont {Smith}, \citenamefont {Lee}, \citenamefont {Edwards}, \citenamefont {Wang}, \citenamefont {Freericks},\ and\ \citenamefont {Monroe}}]{Islam2013}%
  \BibitemOpen
  \bibfield  {author} {\bibinfo {author} {\bibfnamefont {R.}~\bibnamefont {Islam}}, \bibinfo {author} {\bibfnamefont {C.}~\bibnamefont {Senko}}, \bibinfo {author} {\bibfnamefont {W.~C.}\ \bibnamefont {Campbell}}, \bibinfo {author} {\bibfnamefont {S.}~\bibnamefont {Korenblit}}, \bibinfo {author} {\bibfnamefont {J.}~\bibnamefont {Smith}}, \bibinfo {author} {\bibfnamefont {A.}~\bibnamefont {Lee}}, \bibinfo {author} {\bibfnamefont {E.~E.}\ \bibnamefont {Edwards}}, \bibinfo {author} {\bibfnamefont {C.-C.~J.}\ \bibnamefont {Wang}}, \bibinfo {author} {\bibfnamefont {J.~K.}\ \bibnamefont {Freericks}},\ and\ \bibinfo {author} {\bibfnamefont {C.}~\bibnamefont {Monroe}},\ }\bibfield  {title} {\bibinfo {title} {Emergence and frustration of magnetism with variable-range interactions in a quantum simulator},\ }\href {https://doi.org/10.1126/science.1232296} {\bibfield  {journal} {\bibinfo  {journal} {Science}\ }\textbf {\bibinfo {volume} {340}},\ \bibinfo {pages} {583} (\bibinfo {year} {2013})}\BibitemShut {NoStop}%
\bibitem [{\citenamefont {Jurcevic}\ \emph {et~al.}(2014)\citenamefont {Jurcevic}, \citenamefont {Lanyon}, \citenamefont {Hauke}, \citenamefont {Hempel}, \citenamefont {Zoller}, \citenamefont {Blatt},\ and\ \citenamefont {Roos}}]{Jurcevic2014}%
  \BibitemOpen
  \bibfield  {author} {\bibinfo {author} {\bibfnamefont {P.}~\bibnamefont {Jurcevic}}, \bibinfo {author} {\bibfnamefont {B.~P.}\ \bibnamefont {Lanyon}}, \bibinfo {author} {\bibfnamefont {P.}~\bibnamefont {Hauke}}, \bibinfo {author} {\bibfnamefont {C.}~\bibnamefont {Hempel}}, \bibinfo {author} {\bibfnamefont {P.}~\bibnamefont {Zoller}}, \bibinfo {author} {\bibfnamefont {R.}~\bibnamefont {Blatt}},\ and\ \bibinfo {author} {\bibfnamefont {C.~F.}\ \bibnamefont {Roos}},\ }\bibfield  {title} {\bibinfo {title} {Quasiparticle engineering and entanglement propagation in a quantum many-body system},\ }\href {https://doi.org/10.1038/nature13461} {\bibfield  {journal} {\bibinfo  {journal} {Nature}\ }\textbf {\bibinfo {volume} {511}},\ \bibinfo {pages} {202} (\bibinfo {year} {2014})}\BibitemShut {NoStop}%
\bibitem [{\citenamefont {Richerme}\ \emph {et~al.}(2014)\citenamefont {Richerme}, \citenamefont {Gong}, \citenamefont {Lee}, \citenamefont {Senko}, \citenamefont {Smith}, \citenamefont {Foss-Feig}, \citenamefont {Michalakis}, \citenamefont {Gorshkov},\ and\ \citenamefont {Monroe}}]{Richerme2014}%
  \BibitemOpen
  \bibfield  {author} {\bibinfo {author} {\bibfnamefont {P.}~\bibnamefont {Richerme}}, \bibinfo {author} {\bibfnamefont {Z.-X.}\ \bibnamefont {Gong}}, \bibinfo {author} {\bibfnamefont {A.}~\bibnamefont {Lee}}, \bibinfo {author} {\bibfnamefont {C.}~\bibnamefont {Senko}}, \bibinfo {author} {\bibfnamefont {J.}~\bibnamefont {Smith}}, \bibinfo {author} {\bibfnamefont {M.}~\bibnamefont {Foss-Feig}}, \bibinfo {author} {\bibfnamefont {S.}~\bibnamefont {Michalakis}}, \bibinfo {author} {\bibfnamefont {A.~V.}\ \bibnamefont {Gorshkov}},\ and\ \bibinfo {author} {\bibfnamefont {C.}~\bibnamefont {Monroe}},\ }\bibfield  {title} {\bibinfo {title} {Non-local propagation of correlations in quantum systems with long-range interactions},\ }\href {https://doi.org/10.1038/nature13450} {\bibfield  {journal} {\bibinfo  {journal} {Nature}\ }\textbf {\bibinfo {volume} {511}},\ \bibinfo {pages} {198} (\bibinfo {year} {2014})}\BibitemShut {NoStop}%
\bibitem [{\citenamefont {G\"{a}rttner}\ \emph {et~al.}(2017)\citenamefont {G\"{a}rttner}, \citenamefont {Bohnet}, \citenamefont {Safavi-Naini}, \citenamefont {Wall}, \citenamefont {Bollinger},\ and\ \citenamefont {Rey}}]{Grttner2017}%
  \BibitemOpen
  \bibfield  {author} {\bibinfo {author} {\bibfnamefont {M.}~\bibnamefont {G\"{a}rttner}}, \bibinfo {author} {\bibfnamefont {J.~G.}\ \bibnamefont {Bohnet}}, \bibinfo {author} {\bibfnamefont {A.}~\bibnamefont {Safavi-Naini}}, \bibinfo {author} {\bibfnamefont {M.~L.}\ \bibnamefont {Wall}}, \bibinfo {author} {\bibfnamefont {J.~J.}\ \bibnamefont {Bollinger}},\ and\ \bibinfo {author} {\bibfnamefont {A.~M.}\ \bibnamefont {Rey}},\ }\bibfield  {title} {\bibinfo {title} {Measuring out-of-time-order correlations and multiple quantum spectra in a trapped-ion quantum magnet},\ }\href {https://doi.org/10.1038/nphys4119} {\bibfield  {journal} {\bibinfo  {journal} {Nature Physics}\ }\textbf {\bibinfo {volume} {13}},\ \bibinfo {pages} {781} (\bibinfo {year} {2017})}\BibitemShut {NoStop}%
\bibitem [{\citenamefont {Signoles}\ \emph {et~al.}(2021)\citenamefont {Signoles}, \citenamefont {Franz}, \citenamefont {Ferracini~Alves}, \citenamefont {G\"arttner}, \citenamefont {Whitlock}, \citenamefont {Z\"urn},\ and\ \citenamefont {Weidem\"uller}}]{Whitlock2021}%
  \BibitemOpen
  \bibfield  {author} {\bibinfo {author} {\bibfnamefont {A.}~\bibnamefont {Signoles}}, \bibinfo {author} {\bibfnamefont {T.}~\bibnamefont {Franz}}, \bibinfo {author} {\bibfnamefont {R.}~\bibnamefont {Ferracini~Alves}}, \bibinfo {author} {\bibfnamefont {M.}~\bibnamefont {G\"arttner}}, \bibinfo {author} {\bibfnamefont {S.}~\bibnamefont {Whitlock}}, \bibinfo {author} {\bibfnamefont {G.}~\bibnamefont {Z\"urn}},\ and\ \bibinfo {author} {\bibfnamefont {M.}~\bibnamefont {Weidem\"uller}},\ }\bibfield  {title} {\bibinfo {title} {Glassy dynamics in a disordered heisenberg quantum spin system},\ }\href {https://doi.org/10.1103/PhysRevX.11.011011} {\bibfield  {journal} {\bibinfo  {journal} {Phys. Rev. X}\ }\textbf {\bibinfo {volume} {11}},\ \bibinfo {pages} {011011} (\bibinfo {year} {2021})}\BibitemShut {NoStop}%
\bibitem [{\citenamefont {Nauts}\ and\ \citenamefont {Wyatt}(1983)}]{Nauts1983}%
  \BibitemOpen
  \bibfield  {author} {\bibinfo {author} {\bibfnamefont {A.}~\bibnamefont {Nauts}}\ and\ \bibinfo {author} {\bibfnamefont {R.~E.}\ \bibnamefont {Wyatt}},\ }\bibfield  {title} {\bibinfo {title} {New approach to many-state quantum dynamics: The recursive-residue-generation method},\ }\href {https://doi.org/10.1103/PhysRevLett.51.2238} {\bibfield  {journal} {\bibinfo  {journal} {Phys. Rev. Lett.}\ }\textbf {\bibinfo {volume} {51}},\ \bibinfo {pages} {2238} (\bibinfo {year} {1983})}\BibitemShut {NoStop}%
\bibitem [{\citenamefont {Park}\ and\ \citenamefont {Light}(1986)}]{Park1986}%
  \BibitemOpen
  \bibfield  {author} {\bibinfo {author} {\bibfnamefont {T.~J.}\ \bibnamefont {Park}}\ and\ \bibinfo {author} {\bibfnamefont {J.~C.}\ \bibnamefont {Light}},\ }\bibfield  {title} {\bibinfo {title} {{Unitary quantum time evolution by iterative Lanczos reduction}},\ }\href {https://doi.org/10.1063/1.451548} {\bibfield  {journal} {\bibinfo  {journal} {The Journal of Chemical Physics}\ }\textbf {\bibinfo {volume} {85}},\ \bibinfo {pages} {5870} (\bibinfo {year} {1986})}\BibitemShut {NoStop}%
\bibitem [{\citenamefont {Colmenarez}\ and\ \citenamefont {Luitz}(2020)}]{Colmenarez2020}%
  \BibitemOpen
  \bibfield  {author} {\bibinfo {author} {\bibfnamefont {L.}~\bibnamefont {Colmenarez}}\ and\ \bibinfo {author} {\bibfnamefont {D.~J.}\ \bibnamefont {Luitz}},\ }\bibfield  {title} {\bibinfo {title} {Lieb-robinson bounds and out-of-time order correlators in a long-range spin chain},\ }\href {https://doi.org/10.1103/PhysRevResearch.2.043047} {\bibfield  {journal} {\bibinfo  {journal} {Phys. Rev. Res.}\ }\textbf {\bibinfo {volume} {2}},\ \bibinfo {pages} {043047} (\bibinfo {year} {2020})}\BibitemShut {NoStop}%
\bibitem [{\citenamefont {Vahedi}(2022)}]{vahedi2022}%
  \BibitemOpen
  \bibfield  {author} {\bibinfo {author} {\bibfnamefont {J.}~\bibnamefont {Vahedi}},\ }\bibfield  {title} {\bibinfo {title} {{Asymmetric transport in long-range interacting chiral spin chains}},\ }\href {https://doi.org/10.21468/SciPostPhysCore.5.2.021} {\bibfield  {journal} {\bibinfo  {journal} {SciPost Phys. Core}\ }\textbf {\bibinfo {volume} {5}},\ \bibinfo {pages} {021} (\bibinfo {year} {2022})}\BibitemShut {NoStop}%
\bibitem [{\citenamefont {Faridfar}\ and\ \citenamefont {Vahedi}(2022)}]{Faridfar2022}%
  \BibitemOpen
  \bibfield  {author} {\bibinfo {author} {\bibfnamefont {M.}~\bibnamefont {Faridfar}}\ and\ \bibinfo {author} {\bibfnamefont {J.}~\bibnamefont {Vahedi}},\ }\bibfield  {title} {\bibinfo {title} {Thermodynamic behavior of spin-1 heisenberg chain: a comparative study},\ }\href {https://doi.org/10.1007/s10948-021-06086-4} {\bibfield  {journal} {\bibinfo  {journal} {Journal of Superconductivity and Novel Magnetism}\ }\textbf {\bibinfo {volume} {35}},\ \bibinfo {pages} {519} (\bibinfo {year} {2022})}\BibitemShut {NoStop}%
\bibitem [{\citenamefont {Schollwoeck}(2011)}]{schollwoeckDensitymatrixRenormalizationGroup2011}%
  \BibitemOpen
  \bibfield  {author} {\bibinfo {author} {\bibfnamefont {U.}~\bibnamefont {Schollwoeck}},\ }\bibfield  {title} {\bibinfo {title} {The density-matrix renormalization group in the age of matrix product states},\ }\href {https://doi.org/10.1016/j.aop.2010.09.012} {\bibfield  {journal} {\bibinfo  {journal} {Annals of Physics}\ }\textbf {\bibinfo {volume} {326}},\ \bibinfo {pages} {96} (\bibinfo {year} {2011})}\BibitemShut {NoStop}%
\bibitem [{\citenamefont {Vidal}(2003)}]{Vidal2003}%
  \BibitemOpen
  \bibfield  {author} {\bibinfo {author} {\bibfnamefont {G.}~\bibnamefont {Vidal}},\ }\bibfield  {title} {\bibinfo {title} {Efficient classical simulation of slightly entangled quantum computations},\ }\href {https://doi.org/10.1103/PhysRevLett.91.147902} {\bibfield  {journal} {\bibinfo  {journal} {Phys. Rev. Lett.}\ }\textbf {\bibinfo {volume} {91}},\ \bibinfo {pages} {147902} (\bibinfo {year} {2003})}\BibitemShut {NoStop}%
\bibitem [{\citenamefont {White}\ and\ \citenamefont {Feiguin}(2004)}]{White2004}%
  \BibitemOpen
  \bibfield  {author} {\bibinfo {author} {\bibfnamefont {S.~R.}\ \bibnamefont {White}}\ and\ \bibinfo {author} {\bibfnamefont {A.~E.}\ \bibnamefont {Feiguin}},\ }\bibfield  {title} {\bibinfo {title} {Real-time evolution using the density matrix renormalization group},\ }\href {https://doi.org/10.1103/PhysRevLett.93.076401} {\bibfield  {journal} {\bibinfo  {journal} {Phys. Rev. Lett.}\ }\textbf {\bibinfo {volume} {93}},\ \bibinfo {pages} {076401} (\bibinfo {year} {2004})}\BibitemShut {NoStop}%
\bibitem [{\citenamefont {Zaletel}\ \emph {et~al.}(2015)\citenamefont {Zaletel}, \citenamefont {Mong}, \citenamefont {Karrasch}, \citenamefont {Moore},\ and\ \citenamefont {Pollmann}}]{Zaletel2015}%
  \BibitemOpen
  \bibfield  {author} {\bibinfo {author} {\bibfnamefont {M.~P.}\ \bibnamefont {Zaletel}}, \bibinfo {author} {\bibfnamefont {R.~S.~K.}\ \bibnamefont {Mong}}, \bibinfo {author} {\bibfnamefont {C.}~\bibnamefont {Karrasch}}, \bibinfo {author} {\bibfnamefont {J.~E.}\ \bibnamefont {Moore}},\ and\ \bibinfo {author} {\bibfnamefont {F.}~\bibnamefont {Pollmann}},\ }\bibfield  {title} {\bibinfo {title} {Time-evolving a matrix product state with long-ranged interactions},\ }\href {https://doi.org/10.1103/PhysRevB.91.165112} {\bibfield  {journal} {\bibinfo  {journal} {Phys. Rev. B}\ }\textbf {\bibinfo {volume} {91}},\ \bibinfo {pages} {165112} (\bibinfo {year} {2015})}\BibitemShut {NoStop}%
\bibitem [{\citenamefont {Paeckel}\ \emph {et~al.}(2019)\citenamefont {Paeckel}, \citenamefont {Köhler}, \citenamefont {Swoboda}, \citenamefont {Manmana}, \citenamefont {Schollwöck},\ and\ \citenamefont {Hubig}}]{Schollwock2019}%
  \BibitemOpen
  \bibfield  {author} {\bibinfo {author} {\bibfnamefont {S.}~\bibnamefont {Paeckel}}, \bibinfo {author} {\bibfnamefont {T.}~\bibnamefont {Köhler}}, \bibinfo {author} {\bibfnamefont {A.}~\bibnamefont {Swoboda}}, \bibinfo {author} {\bibfnamefont {S.~R.}\ \bibnamefont {Manmana}}, \bibinfo {author} {\bibfnamefont {U.}~\bibnamefont {Schollwöck}},\ and\ \bibinfo {author} {\bibfnamefont {C.}~\bibnamefont {Hubig}},\ }\bibfield  {title} {\bibinfo {title} {Time-evolution methods for matrix-product states},\ }\href {https://doi.org/https://doi.org/10.1016/j.aop.2019.167998} {\bibfield  {journal} {\bibinfo  {journal} {Annals of Physics}\ }\textbf {\bibinfo {volume} {411}},\ \bibinfo {pages} {167998} (\bibinfo {year} {2019})}\BibitemShut {NoStop}%
\bibitem [{\citenamefont {Blakie{\textdagger}}\ \emph {et~al.}(2008)\citenamefont {Blakie{\textdagger}}, \citenamefont {Bradley{\textdagger}}, \citenamefont {Davis}, \citenamefont {Ballagh},\ and\ \citenamefont {Gardiner}}]{Blakie2008}%
  \BibitemOpen
  \bibfield  {author} {\bibinfo {author} {\bibfnamefont {P.}~\bibnamefont {Blakie{\textdagger}}}, \bibinfo {author} {\bibfnamefont {A.}~\bibnamefont {Bradley{\textdagger}}}, \bibinfo {author} {\bibfnamefont {M.}~\bibnamefont {Davis}}, \bibinfo {author} {\bibfnamefont {R.}~\bibnamefont {Ballagh}},\ and\ \bibinfo {author} {\bibfnamefont {C.}~\bibnamefont {Gardiner}},\ }\bibfield  {title} {\bibinfo {title} {Dynamics and statistical mechanics of ultra-cold bose gases using c-field techniques},\ }\href {https://doi.org/10.1080/00018730802564254} {\bibfield  {journal} {\bibinfo  {journal} {Advances in Physics}\ }\textbf {\bibinfo {volume} {57}},\ \bibinfo {pages} {363} (\bibinfo {year} {2008})}\BibitemShut {NoStop}%
\bibitem [{\citenamefont {Polkovnikov}(2010)}]{Polkovnikov2010}%
  \BibitemOpen
  \bibfield  {author} {\bibinfo {author} {\bibfnamefont {A.}~\bibnamefont {Polkovnikov}},\ }\bibfield  {title} {\bibinfo {title} {Phase space representation of quantum dynamics},\ }\href {https://doi.org/10.1016/j.aop.2010.02.006} {\bibfield  {journal} {\bibinfo  {journal} {Annals of Physics}\ }\textbf {\bibinfo {volume} {325}},\ \bibinfo {pages} {1790} (\bibinfo {year} {2010})}\BibitemShut {NoStop}%
\bibitem [{\citenamefont {Tuchman}\ \emph {et~al.}(2006)\citenamefont {Tuchman}, \citenamefont {Orzel}, \citenamefont {Polkovnikov},\ and\ \citenamefont {Kasevich}}]{Tuchman2006}%
  \BibitemOpen
  \bibfield  {author} {\bibinfo {author} {\bibfnamefont {A.~K.}\ \bibnamefont {Tuchman}}, \bibinfo {author} {\bibfnamefont {C.}~\bibnamefont {Orzel}}, \bibinfo {author} {\bibfnamefont {A.}~\bibnamefont {Polkovnikov}},\ and\ \bibinfo {author} {\bibfnamefont {M.~A.}\ \bibnamefont {Kasevich}},\ }\bibfield  {title} {\bibinfo {title} {Nonequilibrium coherence dynamics of a soft boson lattice},\ }\href {https://doi.org/10.1103/PhysRevA.74.051601} {\bibfield  {journal} {\bibinfo  {journal} {Phys. Rev. A}\ }\textbf {\bibinfo {volume} {74}},\ \bibinfo {pages} {051601} (\bibinfo {year} {2006})}\BibitemShut {NoStop}%
\bibitem [{\citenamefont {Davidson}\ \emph {et~al.}(2017)\citenamefont {Davidson}, \citenamefont {Sels},\ and\ \citenamefont {Polkovnikov}}]{Davidson2017}%
  \BibitemOpen
  \bibfield  {author} {\bibinfo {author} {\bibfnamefont {S.~M.}\ \bibnamefont {Davidson}}, \bibinfo {author} {\bibfnamefont {D.}~\bibnamefont {Sels}},\ and\ \bibinfo {author} {\bibfnamefont {A.}~\bibnamefont {Polkovnikov}},\ }\bibfield  {title} {\bibinfo {title} {Semiclassical approach to dynamics of interacting fermions},\ }\href {https://doi.org/10.1016/j.aop.2017.07.003} {\bibfield  {journal} {\bibinfo  {journal} {Annals of Physics}\ }\textbf {\bibinfo {volume} {384}},\ \bibinfo {pages} {128} (\bibinfo {year} {2017})}\BibitemShut {NoStop}%
\bibitem [{\citenamefont {Nagao}\ \emph {et~al.}(2019)\citenamefont {Nagao}, \citenamefont {Kunimi}, \citenamefont {Takasu}, \citenamefont {Takahashi},\ and\ \citenamefont {Danshita}}]{Nagao2019}%
  \BibitemOpen
  \bibfield  {author} {\bibinfo {author} {\bibfnamefont {K.}~\bibnamefont {Nagao}}, \bibinfo {author} {\bibfnamefont {M.}~\bibnamefont {Kunimi}}, \bibinfo {author} {\bibfnamefont {Y.}~\bibnamefont {Takasu}}, \bibinfo {author} {\bibfnamefont {Y.}~\bibnamefont {Takahashi}},\ and\ \bibinfo {author} {\bibfnamefont {I.}~\bibnamefont {Danshita}},\ }\bibfield  {title} {\bibinfo {title} {Semiclassical quench dynamics of bose gases in optical lattices},\ }\href {https://doi.org/10.1103/PhysRevA.99.023622} {\bibfield  {journal} {\bibinfo  {journal} {Phys. Rev. A}\ }\textbf {\bibinfo {volume} {99}},\ \bibinfo {pages} {023622} (\bibinfo {year} {2019})}\BibitemShut {NoStop}%
\bibitem [{\citenamefont {Muleady}\ \emph {et~al.}(2023)\citenamefont {Muleady}, \citenamefont {Yang}, \citenamefont {White},\ and\ \citenamefont {Rey}}]{muleady2023validating}%
  \BibitemOpen
  \bibfield  {author} {\bibinfo {author} {\bibfnamefont {S.~R.}\ \bibnamefont {Muleady}}, \bibinfo {author} {\bibfnamefont {M.}~\bibnamefont {Yang}}, \bibinfo {author} {\bibfnamefont {S.~R.}\ \bibnamefont {White}},\ and\ \bibinfo {author} {\bibfnamefont {A.~M.}\ \bibnamefont {Rey}},\ }\bibfield  {title} {\bibinfo {title} {Validating {{Phase-Space Methods}} with {{Tensor Networks}} in {{Two-Dimensional Spin Models}} with {{Power-Law Interactions}}},\ }\href {https://doi.org/10.1103/PhysRevLett.131.150401} {\bibfield  {journal} {\bibinfo  {journal} {Phys. Rev. Lett.}\ }\textbf {\bibinfo {volume} {131}},\ \bibinfo {pages} {150401} (\bibinfo {year} {2023})}\BibitemShut {NoStop}%
\bibitem [{\citenamefont {Mink}\ \emph {et~al.}(2022)\citenamefont {Mink}, \citenamefont {Petrosyan},\ and\ \citenamefont {Fleischhauer}}]{minkHybridDiscretecontinuousTruncated2022}%
  \BibitemOpen
  \bibfield  {author} {\bibinfo {author} {\bibfnamefont {C.~D.}\ \bibnamefont {Mink}}, \bibinfo {author} {\bibfnamefont {D.}~\bibnamefont {Petrosyan}},\ and\ \bibinfo {author} {\bibfnamefont {M.}~\bibnamefont {Fleischhauer}},\ }\bibfield  {title} {\bibinfo {title} {Hybrid discrete-continuous truncated {{Wigner}} approximation for driven, dissipative spin systems},\ }\href {https://doi.org/10.1103/PhysRevResearch.4.043136} {\bibfield  {journal} {\bibinfo  {journal} {{Phys. Rev. Res.}}\ }\textbf {\bibinfo {volume} {4}},\ \bibinfo {pages} {043136} (\bibinfo {year} {2022})}\BibitemShut {NoStop}%
\bibitem [{\citenamefont {Wootters}(1987)}]{WOOTTERS19871}%
  \BibitemOpen
  \bibfield  {author} {\bibinfo {author} {\bibfnamefont {W.~K.}\ \bibnamefont {Wootters}},\ }\bibfield  {title} {\bibinfo {title} {A wigner-function formulation of finite-state quantum mechanics},\ }\href {https://doi.org/https://doi.org/10.1016/0003-4916(87)90176-X} {\bibfield  {journal} {\bibinfo  {journal} {Annals of Physics}\ }\textbf {\bibinfo {volume} {176}},\ \bibinfo {pages} {1} (\bibinfo {year} {1987})}\BibitemShut {NoStop}%
\bibitem [{\citenamefont {Schachenmayer}\ \emph {et~al.}(2015{\natexlab{a}})\citenamefont {Schachenmayer}, \citenamefont {Pikovski},\ and\ \citenamefont {Rey}}]{Schachenmayer2015}%
  \BibitemOpen
  \bibfield  {author} {\bibinfo {author} {\bibfnamefont {J.}~\bibnamefont {Schachenmayer}}, \bibinfo {author} {\bibfnamefont {A.}~\bibnamefont {Pikovski}},\ and\ \bibinfo {author} {\bibfnamefont {A.~M.}\ \bibnamefont {Rey}},\ }\bibfield  {title} {\bibinfo {title} {Many-body quantum spin dynamics with monte carlo trajectories on a discrete phase space},\ }\href {https://doi.org/10.1103/PhysRevX.5.011022} {\bibfield  {journal} {\bibinfo  {journal} {Phys. Rev. X}\ }\textbf {\bibinfo {volume} {5}},\ \bibinfo {pages} {011022} (\bibinfo {year} {2015}{\natexlab{a}})}\BibitemShut {NoStop}%
\bibitem [{\citenamefont {Schachenmayer}\ \emph {et~al.}(2015{\natexlab{b}})\citenamefont {Schachenmayer}, \citenamefont {Pikovski},\ and\ \citenamefont {Rey}}]{Pikovski2015}%
  \BibitemOpen
  \bibfield  {author} {\bibinfo {author} {\bibfnamefont {J.}~\bibnamefont {Schachenmayer}}, \bibinfo {author} {\bibfnamefont {A.}~\bibnamefont {Pikovski}},\ and\ \bibinfo {author} {\bibfnamefont {A.~M.}\ \bibnamefont {Rey}},\ }\bibfield  {title} {\bibinfo {title} {Dynamics of correlations in two-dimensional quantum spin models with long-range interactions: a phase-space monte-carlo study},\ }\href {https://doi.org/10.1088/1367-2630/17/6/065009} {\bibfield  {journal} {\bibinfo  {journal} {New Journal of Physics}\ }\textbf {\bibinfo {volume} {17}},\ \bibinfo {pages} {065009} (\bibinfo {year} {2015}{\natexlab{b}})}\BibitemShut {NoStop}%
\bibitem [{\citenamefont {Pucci}\ \emph {et~al.}(2016)\citenamefont {Pucci}, \citenamefont {Roy},\ and\ \citenamefont {Kastner}}]{Pucci2016}%
  \BibitemOpen
  \bibfield  {author} {\bibinfo {author} {\bibfnamefont {L.}~\bibnamefont {Pucci}}, \bibinfo {author} {\bibfnamefont {A.}~\bibnamefont {Roy}},\ and\ \bibinfo {author} {\bibfnamefont {M.}~\bibnamefont {Kastner}},\ }\bibfield  {title} {\bibinfo {title} {Simulation of quantum spin dynamics by phase space sampling of bogoliubov-born-green-kirkwood-yvon trajectories},\ }\href {https://doi.org/10.1103/PhysRevB.93.174302} {\bibfield  {journal} {\bibinfo  {journal} {Phys. Rev. B}\ }\textbf {\bibinfo {volume} {93}},\ \bibinfo {pages} {174302} (\bibinfo {year} {2016})}\BibitemShut {NoStop}%
\bibitem [{\citenamefont {Acevedo}\ \emph {et~al.}(2017)\citenamefont {Acevedo}, \citenamefont {Safavi-Naini}, \citenamefont {Schachenmayer}, \citenamefont {Wall}, \citenamefont {Nandkishore},\ and\ \citenamefont {Rey}}]{Acevedo2017}%
  \BibitemOpen
  \bibfield  {author} {\bibinfo {author} {\bibfnamefont {O.~L.}\ \bibnamefont {Acevedo}}, \bibinfo {author} {\bibfnamefont {A.}~\bibnamefont {Safavi-Naini}}, \bibinfo {author} {\bibfnamefont {J.}~\bibnamefont {Schachenmayer}}, \bibinfo {author} {\bibfnamefont {M.~L.}\ \bibnamefont {Wall}}, \bibinfo {author} {\bibfnamefont {R.}~\bibnamefont {Nandkishore}},\ and\ \bibinfo {author} {\bibfnamefont {A.~M.}\ \bibnamefont {Rey}},\ }\bibfield  {title} {\bibinfo {title} {Exploring many-body localization and thermalization using semiclassical methods},\ }\href {https://doi.org/10.1103/PhysRevA.96.033604} {\bibfield  {journal} {\bibinfo  {journal} {Phys. Rev. A}\ }\textbf {\bibinfo {volume} {96}},\ \bibinfo {pages} {033604} (\bibinfo {year} {2017})}\BibitemShut {NoStop}%
\bibitem [{\citenamefont {Pi\~neiro Orioli}\ \emph {et~al.}(2017)\citenamefont {Pi\~neiro Orioli}, \citenamefont {Safavi-Naini}, \citenamefont {Wall},\ and\ \citenamefont {Rey}}]{Arghavan2017}%
  \BibitemOpen
  \bibfield  {author} {\bibinfo {author} {\bibfnamefont {A.}~\bibnamefont {Pi\~neiro Orioli}}, \bibinfo {author} {\bibfnamefont {A.}~\bibnamefont {Safavi-Naini}}, \bibinfo {author} {\bibfnamefont {M.~L.}\ \bibnamefont {Wall}},\ and\ \bibinfo {author} {\bibfnamefont {A.~M.}\ \bibnamefont {Rey}},\ }\bibfield  {title} {\bibinfo {title} {Nonequilibrium dynamics of spin-boson models from phase-space methods},\ }\href {https://doi.org/10.1103/PhysRevA.96.033607} {\bibfield  {journal} {\bibinfo  {journal} {Phys. Rev. A}\ }\textbf {\bibinfo {volume} {96}},\ \bibinfo {pages} {033607} (\bibinfo {year} {2017})}\BibitemShut {NoStop}%
\bibitem [{\citenamefont {Czischek}\ \emph {et~al.}(2018)\citenamefont {Czischek}, \citenamefont {G\"{a}rttner}, \citenamefont {Oberthaler}, \citenamefont {Kastner},\ and\ \citenamefont {Gasenzer}}]{Czischek2018}%
  \BibitemOpen
  \bibfield  {author} {\bibinfo {author} {\bibfnamefont {S.}~\bibnamefont {Czischek}}, \bibinfo {author} {\bibfnamefont {M.}~\bibnamefont {G\"{a}rttner}}, \bibinfo {author} {\bibfnamefont {M.}~\bibnamefont {Oberthaler}}, \bibinfo {author} {\bibfnamefont {M.}~\bibnamefont {Kastner}},\ and\ \bibinfo {author} {\bibfnamefont {T.}~\bibnamefont {Gasenzer}},\ }\bibfield  {title} {\bibinfo {title} {Quenches near criticality of the quantum ising chain{\textemdash}power and limitations of the discrete truncated wigner approximation},\ }\href {https://doi.org/10.1088/2058-9565/aae3f7} {\bibfield  {journal} {\bibinfo  {journal} {Quantum Science and Technology}\ }\textbf {\bibinfo {volume} {4}},\ \bibinfo {pages} {014006} (\bibinfo {year} {2018})}\BibitemShut {NoStop}%
\bibitem [{\citenamefont {Pi\~neiro Orioli}\ \emph {et~al.}(2018)\citenamefont {Pi\~neiro Orioli}, \citenamefont {Signoles}, \citenamefont {Wildhagen}, \citenamefont {G\"unter}, \citenamefont {Berges}, \citenamefont {Whitlock},\ and\ \citenamefont {Weidem\"uller}}]{Berges2018}%
  \BibitemOpen
  \bibfield  {author} {\bibinfo {author} {\bibfnamefont {A.}~\bibnamefont {Pi\~neiro Orioli}}, \bibinfo {author} {\bibfnamefont {A.}~\bibnamefont {Signoles}}, \bibinfo {author} {\bibfnamefont {H.}~\bibnamefont {Wildhagen}}, \bibinfo {author} {\bibfnamefont {G.}~\bibnamefont {G\"unter}}, \bibinfo {author} {\bibfnamefont {J.}~\bibnamefont {Berges}}, \bibinfo {author} {\bibfnamefont {S.}~\bibnamefont {Whitlock}},\ and\ \bibinfo {author} {\bibfnamefont {M.}~\bibnamefont {Weidem\"uller}},\ }\bibfield  {title} {\bibinfo {title} {Relaxation of an isolated dipolar-interacting rydberg quantum spin system},\ }\href {https://doi.org/10.1103/PhysRevLett.120.063601} {\bibfield  {journal} {\bibinfo  {journal} {Phys. Rev. Lett.}\ }\textbf {\bibinfo {volume} {120}},\ \bibinfo {pages} {063601} (\bibinfo {year} {2018})}\BibitemShut {NoStop}%
\bibitem [{\citenamefont {Sundar}\ \emph {et~al.}(2019)\citenamefont {Sundar}, \citenamefont {Wang},\ and\ \citenamefont {Hazzard}}]{Bhuvanesh2019}%
  \BibitemOpen
  \bibfield  {author} {\bibinfo {author} {\bibfnamefont {B.}~\bibnamefont {Sundar}}, \bibinfo {author} {\bibfnamefont {K.~C.}\ \bibnamefont {Wang}},\ and\ \bibinfo {author} {\bibfnamefont {K.~R.~A.}\ \bibnamefont {Hazzard}},\ }\bibfield  {title} {\bibinfo {title} {Analysis of continuous and discrete wigner approximations for spin dynamics},\ }\href {https://doi.org/10.1103/PhysRevA.99.043627} {\bibfield  {journal} {\bibinfo  {journal} {Phys. Rev. A}\ }\textbf {\bibinfo {volume} {99}},\ \bibinfo {pages} {043627} (\bibinfo {year} {2019})}\BibitemShut {NoStop}%
\bibitem [{\citenamefont {Khasseh}\ \emph {et~al.}(2020)\citenamefont {Khasseh}, \citenamefont {Russomanno}, \citenamefont {Schmitt}, \citenamefont {Heyl},\ and\ \citenamefont {Fazio}}]{Khasseh2020}%
  \BibitemOpen
  \bibfield  {author} {\bibinfo {author} {\bibfnamefont {R.}~\bibnamefont {Khasseh}}, \bibinfo {author} {\bibfnamefont {A.}~\bibnamefont {Russomanno}}, \bibinfo {author} {\bibfnamefont {M.}~\bibnamefont {Schmitt}}, \bibinfo {author} {\bibfnamefont {M.}~\bibnamefont {Heyl}},\ and\ \bibinfo {author} {\bibfnamefont {R.}~\bibnamefont {Fazio}},\ }\bibfield  {title} {\bibinfo {title} {Discrete truncated wigner approach to dynamical phase transitions in ising models after a quantum quench},\ }\href {https://doi.org/10.1103/PhysRevB.102.014303} {\bibfield  {journal} {\bibinfo  {journal} {Phys. Rev. B}\ }\textbf {\bibinfo {volume} {102}},\ \bibinfo {pages} {014303} (\bibinfo {year} {2020})}\BibitemShut {NoStop}%
\bibitem [{\citenamefont {Kunimi}\ \emph {et~al.}(2021)\citenamefont {Kunimi}, \citenamefont {Nagao}, \citenamefont {Goto},\ and\ \citenamefont {Danshita}}]{Masaya2021}%
  \BibitemOpen
  \bibfield  {author} {\bibinfo {author} {\bibfnamefont {M.}~\bibnamefont {Kunimi}}, \bibinfo {author} {\bibfnamefont {K.}~\bibnamefont {Nagao}}, \bibinfo {author} {\bibfnamefont {S.}~\bibnamefont {Goto}},\ and\ \bibinfo {author} {\bibfnamefont {I.}~\bibnamefont {Danshita}},\ }\bibfield  {title} {\bibinfo {title} {Performance evaluation of the discrete truncated wigner approximation for quench dynamics of quantum spin systems with long-range interactions},\ }\href {https://doi.org/10.1103/PhysRevResearch.3.013060} {\bibfield  {journal} {\bibinfo  {journal} {Phys. Rev. Res.}\ }\textbf {\bibinfo {volume} {3}},\ \bibinfo {pages} {013060} (\bibinfo {year} {2021})}\BibitemShut {NoStop}%
\bibitem [{\citenamefont {Morong}\ \emph {et~al.}(2021)\citenamefont {Morong}, \citenamefont {Muleady}, \citenamefont {Kimchi}, \citenamefont {Xu}, \citenamefont {Nandkishore}, \citenamefont {Rey},\ and\ \citenamefont {DeMarco}}]{Morong2021}%
  \BibitemOpen
  \bibfield  {author} {\bibinfo {author} {\bibfnamefont {W.}~\bibnamefont {Morong}}, \bibinfo {author} {\bibfnamefont {S.~R.}\ \bibnamefont {Muleady}}, \bibinfo {author} {\bibfnamefont {I.}~\bibnamefont {Kimchi}}, \bibinfo {author} {\bibfnamefont {W.}~\bibnamefont {Xu}}, \bibinfo {author} {\bibfnamefont {R.~M.}\ \bibnamefont {Nandkishore}}, \bibinfo {author} {\bibfnamefont {A.~M.}\ \bibnamefont {Rey}},\ and\ \bibinfo {author} {\bibfnamefont {B.}~\bibnamefont {DeMarco}},\ }\bibfield  {title} {\bibinfo {title} {Disorder-controlled relaxation in a three-dimensional hubbard model quantum simulator},\ }\href {https://doi.org/10.1103/PhysRevResearch.3.L012009} {\bibfield  {journal} {\bibinfo  {journal} {Phys. Rev. Res.}\ }\textbf {\bibinfo {volume} {3}},\ \bibinfo {pages} {L012009} (\bibinfo {year} {2021})}\BibitemShut {NoStop}%
\bibitem [{\citenamefont {Davidson}\ and\ \citenamefont {Polkovnikov}(2015)}]{Davidson2015}%
  \BibitemOpen
  \bibfield  {author} {\bibinfo {author} {\bibfnamefont {S.~M.}\ \bibnamefont {Davidson}}\ and\ \bibinfo {author} {\bibfnamefont {A.}~\bibnamefont {Polkovnikov}},\ }\bibfield  {title} {\bibinfo {title} {$su(3)$ semiclassical representation of quantum dynamics of interacting spins},\ }\href {https://doi.org/10.1103/PhysRevLett.114.045701} {\bibfield  {journal} {\bibinfo  {journal} {Phys. Rev. Lett.}\ }\textbf {\bibinfo {volume} {114}},\ \bibinfo {pages} {045701} (\bibinfo {year} {2015})}\BibitemShut {NoStop}%
\bibitem [{\citenamefont {Wurtz}\ \emph {et~al.}(2018)\citenamefont {Wurtz}, \citenamefont {Polkovnikov},\ and\ \citenamefont {Sels}}]{Wurtz2018}%
  \BibitemOpen
  \bibfield  {author} {\bibinfo {author} {\bibfnamefont {J.}~\bibnamefont {Wurtz}}, \bibinfo {author} {\bibfnamefont {A.}~\bibnamefont {Polkovnikov}},\ and\ \bibinfo {author} {\bibfnamefont {D.}~\bibnamefont {Sels}},\ }\bibfield  {title} {\bibinfo {title} {Cluster truncated wigner approximation in strongly interacting systems},\ }\href {https://doi.org/10.1016/j.aop.2018.06.001} {\bibfield  {journal} {\bibinfo  {journal} {Annals of Physics}\ }\textbf {\bibinfo {volume} {395}},\ \bibinfo {pages} {341} (\bibinfo {year} {2018})}\BibitemShut {NoStop}%
\bibitem [{\citenamefont {Zhu}\ \emph {et~al.}(2019)\citenamefont {Zhu}, \citenamefont {Rey},\ and\ \citenamefont {Schachenmayer}}]{Zhu2019}%
  \BibitemOpen
  \bibfield  {author} {\bibinfo {author} {\bibfnamefont {B.}~\bibnamefont {Zhu}}, \bibinfo {author} {\bibfnamefont {A.~M.}\ \bibnamefont {Rey}},\ and\ \bibinfo {author} {\bibfnamefont {J.}~\bibnamefont {Schachenmayer}},\ }\bibfield  {title} {\bibinfo {title} {A generalized phase space approach for solving quantum spin dynamics},\ }\href {https://doi.org/10.1088/1367-2630/ab354d} {\bibfield  {journal} {\bibinfo  {journal} {New Journal of Physics}\ }\textbf {\bibinfo {volume} {21}},\ \bibinfo {pages} {082001} (\bibinfo {year} {2019})}\BibitemShut {NoStop}%
\bibitem [{\citenamefont {Nagao}\ and\ \citenamefont {Yunoki}(2024)}]{nagaoTwodimensionalCorrelationPropagation2024}%
  \BibitemOpen
  \bibfield  {author} {\bibinfo {author} {\bibfnamefont {K.}~\bibnamefont {Nagao}}\ and\ \bibinfo {author} {\bibfnamefont {S.}~\bibnamefont {Yunoki}},\ }\href@noop {} {\bibinfo {title} {Two-dimensional correlation propagation dynamics with a cluster discrete phase-space method}} (\bibinfo {year} {2024}),\ \Eprint {https://arxiv.org/abs/2404.18594} {arxiv:2404.18594 [cond-mat, physics:quant-ph]} \BibitemShut {NoStop}%
\bibitem [{\citenamefont {Mohdeb}\ \emph {et~al.}(2020)\citenamefont {Mohdeb}, \citenamefont {Vahedi}, \citenamefont {Moure}, \citenamefont {Roshani}, \citenamefont {Lee}, \citenamefont {Bhatt}, \citenamefont {Kettemann},\ and\ \citenamefont {Haas}}]{Mohdeb2020}%
  \BibitemOpen
  \bibfield  {author} {\bibinfo {author} {\bibfnamefont {Y.}~\bibnamefont {Mohdeb}}, \bibinfo {author} {\bibfnamefont {J.}~\bibnamefont {Vahedi}}, \bibinfo {author} {\bibfnamefont {N.}~\bibnamefont {Moure}}, \bibinfo {author} {\bibfnamefont {A.}~\bibnamefont {Roshani}}, \bibinfo {author} {\bibfnamefont {H.-Y.}\ \bibnamefont {Lee}}, \bibinfo {author} {\bibfnamefont {R.~N.}\ \bibnamefont {Bhatt}}, \bibinfo {author} {\bibfnamefont {S.}~\bibnamefont {Kettemann}},\ and\ \bibinfo {author} {\bibfnamefont {S.}~\bibnamefont {Haas}},\ }\bibfield  {title} {\bibinfo {title} {Entanglement properties of disordered quantum spin chains with long-range antiferromagnetic interactions},\ }\href {https://doi.org/10.1103/PhysRevB.102.214201} {\bibfield  {journal} {\bibinfo  {journal} {Phys. Rev. B}\ }\textbf {\bibinfo {volume} {102}},\ \bibinfo {pages} {214201} (\bibinfo {year} {2020})}\BibitemShut {NoStop}%
\bibitem [{\citenamefont {Mohdeb}\ \emph {et~al.}(2022)\citenamefont {Mohdeb}, \citenamefont {Vahedi},\ and\ \citenamefont {Kettemann}}]{Mohdeb2022}%
  \BibitemOpen
  \bibfield  {author} {\bibinfo {author} {\bibfnamefont {Y.}~\bibnamefont {Mohdeb}}, \bibinfo {author} {\bibfnamefont {J.}~\bibnamefont {Vahedi}},\ and\ \bibinfo {author} {\bibfnamefont {S.}~\bibnamefont {Kettemann}},\ }\bibfield  {title} {\bibinfo {title} {Excited-eigenstate entanglement properties of xx spin chains with random long-range interactions},\ }\href {https://doi.org/10.1103/PhysRevB.106.104201} {\bibfield  {journal} {\bibinfo  {journal} {Phys. Rev. B}\ }\textbf {\bibinfo {volume} {106}},\ \bibinfo {pages} {104201} (\bibinfo {year} {2022})}\BibitemShut {NoStop}%
\bibitem [{\citenamefont {Mohdeb}\ \emph {et~al.}(2023)\citenamefont {Mohdeb}, \citenamefont {Vahedi}, \citenamefont {Bhatt}, \citenamefont {Haas},\ and\ \citenamefont {Kettemann}}]{Mohdeb2023}%
  \BibitemOpen
  \bibfield  {author} {\bibinfo {author} {\bibfnamefont {Y.}~\bibnamefont {Mohdeb}}, \bibinfo {author} {\bibfnamefont {J.}~\bibnamefont {Vahedi}}, \bibinfo {author} {\bibfnamefont {R.~N.}\ \bibnamefont {Bhatt}}, \bibinfo {author} {\bibfnamefont {S.}~\bibnamefont {Haas}},\ and\ \bibinfo {author} {\bibfnamefont {S.}~\bibnamefont {Kettemann}},\ }\bibfield  {title} {\bibinfo {title} {Global quench dynamics and the growth of entanglement entropy in disordered spin chains with tunable range interactions},\ }\href {https://doi.org/10.1103/PhysRevB.108.L140203} {\bibfield  {journal} {\bibinfo  {journal} {Phys. Rev. B}\ }\textbf {\bibinfo {volume} {108}},\ \bibinfo {pages} {L140203} (\bibinfo {year} {2023})}\BibitemShut {NoStop}%
\bibitem [{\citenamefont {Altman}\ and\ \citenamefont {Vosk}(2015)}]{altmanUniversalDynamicsRenormalization2015}%
  \BibitemOpen
  \bibfield  {author} {\bibinfo {author} {\bibfnamefont {E.}~\bibnamefont {Altman}}\ and\ \bibinfo {author} {\bibfnamefont {R.}~\bibnamefont {Vosk}},\ }\bibfield  {title} {\bibinfo {title} {Universal {{Dynamics}} and {{Renormalization}} in {{Many-Body-Localized Systems}}},\ }\href {https://doi.org/10.1146/annurev-conmatphys-031214-014701} {\bibfield  {journal} {\bibinfo  {journal} {Annual Review of Condensed Matter Physics}\ }\textbf {\bibinfo {volume} {6}},\ \bibinfo {pages} {383} (\bibinfo {year} {2015})}\BibitemShut {NoStop}%
\bibitem [{\citenamefont {Vasseur}\ \emph {et~al.}(2015)\citenamefont {Vasseur}, \citenamefont {Potter},\ and\ \citenamefont {Parameswaran}}]{vasseurQuantumCriticalityHot2015}%
  \BibitemOpen
  \bibfield  {author} {\bibinfo {author} {\bibfnamefont {R.}~\bibnamefont {Vasseur}}, \bibinfo {author} {\bibfnamefont {A.~C.}\ \bibnamefont {Potter}},\ and\ \bibinfo {author} {\bibfnamefont {S.~A.}\ \bibnamefont {Parameswaran}},\ }\bibfield  {title} {\bibinfo {title} {Quantum {{Criticality}} of {{Hot Random Spin Chains}}},\ }\href {https://doi.org/10.1103/PhysRevLett.114.217201} {\bibfield  {journal} {\bibinfo  {journal} {Physical Review Letters}\ }\textbf {\bibinfo {volume} {114}},\ \bibinfo {pages} {217201} (\bibinfo {year} {2015})}\BibitemShut {NoStop}%
\bibitem [{\citenamefont {Igl{\'o}i}\ and\ \citenamefont {Monthus}(2018)}]{igloiStrongDisorderRG2018}%
  \BibitemOpen
  \bibfield  {author} {\bibinfo {author} {\bibfnamefont {F.}~\bibnamefont {Igl{\'o}i}}\ and\ \bibinfo {author} {\bibfnamefont {C.}~\bibnamefont {Monthus}},\ }\bibfield  {title} {\bibinfo {title} {Strong disorder {{RG}} approach -- a short review of recent developments},\ }\href {https://doi.org/10.1140/epjb/e2018-90434-8} {\bibfield  {journal} {\bibinfo  {journal} {The European Physical Journal B}\ }\textbf {\bibinfo {volume} {91}},\ \bibinfo {pages} {290} (\bibinfo {year} {2018})}\BibitemShut {NoStop}%
\bibitem [{Note1()}]{Note1}%
  \BibitemOpen
  \bibinfo {note} {\protect \href {https://github.com/abraemer/DiscreteCTWAPaper}{https://github.com/abraemer/DiscreteCTWAPaper}}\BibitemShut {NoStop}%
\bibitem [{\citenamefont {Braemer}\ \emph {et~al.}(2022)\citenamefont {Braemer}, \citenamefont {Franz}, \citenamefont {Weidem\"uller},\ and\ \citenamefont {G\"arttner}}]{Braemer2022}%
  \BibitemOpen
  \bibfield  {author} {\bibinfo {author} {\bibfnamefont {A.}~\bibnamefont {Braemer}}, \bibinfo {author} {\bibfnamefont {T.}~\bibnamefont {Franz}}, \bibinfo {author} {\bibfnamefont {M.}~\bibnamefont {Weidem\"uller}},\ and\ \bibinfo {author} {\bibfnamefont {M.}~\bibnamefont {G\"arttner}},\ }\bibfield  {title} {\bibinfo {title} {Pair localization in dipolar systems with tunable positional disorder},\ }\href {https://doi.org/10.1103/PhysRevB.106.134212} {\bibfield  {journal} {\bibinfo  {journal} {Phys. Rev. B}\ }\textbf {\bibinfo {volume} {106}},\ \bibinfo {pages} {134212} (\bibinfo {year} {2022})}\BibitemShut {NoStop}%
\bibitem [{\citenamefont {Geier}\ \emph {et~al.}(2021)\citenamefont {Geier}, \citenamefont {Thaicharoen}, \citenamefont {Hainaut}, \citenamefont {Franz}, \citenamefont {Salzinger}, \citenamefont {Tebben}, \citenamefont {Grimshandl}, \citenamefont {Z{\"u}rn},\ and\ \citenamefont {Weidem{\"u}ller}}]{geierFloquetHamiltonianEngineering2021}%
  \BibitemOpen
  \bibfield  {author} {\bibinfo {author} {\bibfnamefont {S.}~\bibnamefont {Geier}}, \bibinfo {author} {\bibfnamefont {N.}~\bibnamefont {Thaicharoen}}, \bibinfo {author} {\bibfnamefont {C.}~\bibnamefont {Hainaut}}, \bibinfo {author} {\bibfnamefont {T.}~\bibnamefont {Franz}}, \bibinfo {author} {\bibfnamefont {A.}~\bibnamefont {Salzinger}}, \bibinfo {author} {\bibfnamefont {A.}~\bibnamefont {Tebben}}, \bibinfo {author} {\bibfnamefont {D.}~\bibnamefont {Grimshandl}}, \bibinfo {author} {\bibfnamefont {G.}~\bibnamefont {Z{\"u}rn}},\ and\ \bibinfo {author} {\bibfnamefont {M.}~\bibnamefont {Weidem{\"u}ller}},\ }\bibfield  {title} {\bibinfo {title} {Floquet {{Hamiltonian}} engineering of an isolated many-body spin system},\ }\href {https://doi.org/10.1126/science.abd9547} {\bibfield  {journal} {\bibinfo  {journal} {Science}\ }\textbf {\bibinfo {volume} {374}},\ \bibinfo {pages} {1149} (\bibinfo {year} {2021})}\BibitemShut {NoStop}%
\bibitem [{\citenamefont {Bezanson}\ \emph {et~al.}(2017)\citenamefont {Bezanson}, \citenamefont {Edelman}, \citenamefont {Karpinski},\ and\ \citenamefont {Shah}}]{bezansonJuliaFreshApproach2017}%
  \BibitemOpen
  \bibfield  {author} {\bibinfo {author} {\bibfnamefont {J.}~\bibnamefont {Bezanson}}, \bibinfo {author} {\bibfnamefont {A.}~\bibnamefont {Edelman}}, \bibinfo {author} {\bibfnamefont {S.}~\bibnamefont {Karpinski}},\ and\ \bibinfo {author} {\bibfnamefont {V.~B.}\ \bibnamefont {Shah}},\ }\bibfield  {title} {\bibinfo {title} {Julia: {{A Fresh Approach}} to {{Numerical Computing}}},\ }\href {https://doi.org/10.1137/141000671} {\bibfield  {journal} {\bibinfo  {journal} {SIAM Review}\ }\textbf {\bibinfo {volume} {59}},\ \bibinfo {pages} {65} (\bibinfo {year} {2017})}\BibitemShut {NoStop}%
\bibitem [{\citenamefont {Datseris}\ \emph {et~al.}(2020)\citenamefont {Datseris}, \citenamefont {Isensee}, \citenamefont {Pech},\ and\ \citenamefont {G{\'a}l}}]{datserisDrWatsonPerfectSidekick2020}%
  \BibitemOpen
  \bibfield  {author} {\bibinfo {author} {\bibfnamefont {G.}~\bibnamefont {Datseris}}, \bibinfo {author} {\bibfnamefont {J.}~\bibnamefont {Isensee}}, \bibinfo {author} {\bibfnamefont {S.}~\bibnamefont {Pech}},\ and\ \bibinfo {author} {\bibfnamefont {T.}~\bibnamefont {G{\'a}l}},\ }\bibfield  {title} {\bibinfo {title} {{{DrWatson}}: The perfect sidekick for your scientific inquiries},\ }\href {https://doi.org/10.21105/joss.02673} {\bibfield  {journal} {\bibinfo  {journal} {Journal of Open Source Software}\ }\textbf {\bibinfo {volume} {5}},\ \bibinfo {pages} {2673} (\bibinfo {year} {2020})}\BibitemShut {NoStop}%
\bibitem [{\citenamefont {{Fons van der Plas}}\ \emph {et~al.}(2024)\citenamefont {{Fons van der Plas}}, \citenamefont {Dral}, \citenamefont {Berg}, \citenamefont {{$\Gamma\varepsilon\omega\rho\gamma\alpha\kappa$}{\'o}{$\pi$}o{$\upsilon\lambda$}o{$\varsigma$}}, \citenamefont {Huijzer}, \citenamefont {Boche{\'n}ski}, \citenamefont {Mengali}, \citenamefont {Burns}, \citenamefont {Lungwitz}, \citenamefont {Priyashan}, \citenamefont {Ling}, \citenamefont {Wu}, \citenamefont {Kadowaki}, \citenamefont {Zhang}, \citenamefont {Schneider}, \citenamefont {Weaver}, \citenamefont {{Xiu-zhe (Roger) Luo}}, \citenamefont {Gerritsen}, \citenamefont {Novosel}, \citenamefont {Supanat}, \citenamefont {Moon}, \citenamefont {M{\"u}ller}, \citenamefont {Timothy}, \citenamefont {Flore}, \citenamefont {Jeremiah}, \citenamefont {O'Mara}, \citenamefont {Hatherly},\ and\ \citenamefont {{kcin96}}}]{fonsvanderplasFonspPlutoJl2024}%
  \BibitemOpen
  \bibfield  {author} {\bibinfo {author} {\bibnamefont {{Fons van der Plas}}}, \bibinfo {author} {\bibfnamefont {M.}~\bibnamefont {Dral}}, \bibinfo {author} {\bibfnamefont {P.}~\bibnamefont {Berg}}, \bibinfo {author} {\bibfnamefont {P.}~\bibnamefont {{$\Gamma\varepsilon\omega\rho\gamma\alpha\kappa$}{\'o}{$\pi$}o{$\upsilon\lambda$}o{$\varsigma$}}}, \bibinfo {author} {\bibfnamefont {R.}~\bibnamefont {Huijzer}}, \bibinfo {author} {\bibfnamefont {M.}~\bibnamefont {Boche{\'n}ski}}, \bibinfo {author} {\bibfnamefont {A.}~\bibnamefont {Mengali}}, \bibinfo {author} {\bibfnamefont {C.}~\bibnamefont {Burns}}, \bibinfo {author} {\bibfnamefont {B.}~\bibnamefont {Lungwitz}}, \bibinfo {author} {\bibfnamefont {H.}~\bibnamefont {Priyashan}}, \bibinfo {author} {\bibfnamefont {J.}~\bibnamefont {Ling}}, \bibinfo {author} {\bibfnamefont {G.}~\bibnamefont {Wu}}, \bibinfo {author} {\bibfnamefont {S.}~\bibnamefont {Kadowaki}}, \bibinfo {author} {\bibfnamefont {E.}~\bibnamefont {Zhang}}, \bibinfo {author} {\bibfnamefont {F.~S.~S.}\
  \bibnamefont {Schneider}}, \bibinfo {author} {\bibfnamefont {I.}~\bibnamefont {Weaver}}, \bibinfo {author} {\bibnamefont {{Xiu-zhe (Roger) Luo}}}, \bibinfo {author} {\bibfnamefont {J.}~\bibnamefont {Gerritsen}}, \bibinfo {author} {\bibfnamefont {R.}~\bibnamefont {Novosel}}, \bibinfo {author} {\bibnamefont {Supanat}}, \bibinfo {author} {\bibfnamefont {Z.}~\bibnamefont {Moon}}, \bibinfo {author} {\bibfnamefont {L.}~\bibnamefont {M{\"u}ller}}, \bibinfo {author} {\bibnamefont {Timothy}}, \bibinfo {author} {\bibfnamefont {V.}~\bibnamefont {Flore}}, \bibinfo {author} {\bibnamefont {Jeremiah}}, \bibinfo {author} {\bibfnamefont {C.}~\bibnamefont {O'Mara}}, \bibinfo {author} {\bibfnamefont {M.}~\bibnamefont {Hatherly}},\ and\ \bibinfo {author} {\bibnamefont {{kcin96}}},\ }\href {https://doi.org/10.5281/ZENODO.4792401} {\bibinfo {title} {Fonsp/{{Pluto}}.jl: V0.19.42}},\ \bibinfo {howpublished} {Zenodo} (\bibinfo {year} {2024})\BibitemShut {NoStop}%
\bibitem [{\citenamefont {Rackauckas}\ and\ \citenamefont {Nie}(2017)}]{rackauckas2017differentialequations}%
  \BibitemOpen
  \bibfield  {author} {\bibinfo {author} {\bibfnamefont {C.}~\bibnamefont {Rackauckas}}\ and\ \bibinfo {author} {\bibfnamefont {Q.}~\bibnamefont {Nie}},\ }\bibfield  {title} {\bibinfo {title} {Differentialequations.jl--a performant and feature-rich ecosystem for solving differential equations in julia},\ }\href {https://doi.org/10.5334/jors.151} {\bibfield  {journal} {\bibinfo  {journal} {Journal of Open Research Software}\ }\textbf {\bibinfo {volume} {5}},\ \bibinfo {pages} {15} (\bibinfo {year} {2017})}\BibitemShut {NoStop}%
\bibitem [{\citenamefont {Verner}(2010)}]{vernerNumericallyOptimalRunge2010}%
  \BibitemOpen
  \bibfield  {author} {\bibinfo {author} {\bibfnamefont {J.~H.}\ \bibnamefont {Verner}},\ }\bibfield  {title} {\bibinfo {title} {Numerically optimal {{Runge}}--{{Kutta}} pairs with interpolants},\ }\href {https://doi.org/10.1007/s11075-009-9290-3} {\bibfield  {journal} {\bibinfo  {journal} {Numerical Algorithms}\ }\textbf {\bibinfo {volume} {53}},\ \bibinfo {pages} {383} (\bibinfo {year} {2010})}\BibitemShut {NoStop}%
\bibitem [{\citenamefont {Danisch}\ and\ \citenamefont {Krumbiegel}(2021)}]{danischMakieJlFlexible2021}%
  \BibitemOpen
  \bibfield  {author} {\bibinfo {author} {\bibfnamefont {S.}~\bibnamefont {Danisch}}\ and\ \bibinfo {author} {\bibfnamefont {J.}~\bibnamefont {Krumbiegel}},\ }\bibfield  {title} {\bibinfo {title} {Makie.jl: {{Flexible}} high-performance data visualization for {{Julia}}},\ }\href {https://doi.org/10.21105/joss.03349} {\bibfield  {journal} {\bibinfo  {journal} {Journal of Open Source Software}\ }\textbf {\bibinfo {volume} {6}},\ \bibinfo {pages} {3349} (\bibinfo {year} {2021})}\BibitemShut {NoStop}%
\bibitem [{\citenamefont {{Bouchet-Valat}}\ and\ \citenamefont {Kami{\'n}ski}(2023)}]{bouchet-valatDataframesJlFlexible2023}%
  \BibitemOpen
  \bibfield  {author} {\bibinfo {author} {\bibfnamefont {M.}~\bibnamefont {{Bouchet-Valat}}}\ and\ \bibinfo {author} {\bibfnamefont {B.}~\bibnamefont {Kami{\'n}ski}},\ }\bibfield  {title} {\bibinfo {title} {Dataframes.jl: {{Flexible}} and fast tabular data in julia},\ }\href {https://doi.org/10.18637/jss.v107.i04} {\bibfield  {journal} {\bibinfo  {journal} {Journal of Statistical Software}\ }\textbf {\bibinfo {volume} {107}},\ \bibinfo {pages} {1} (\bibinfo {year} {2023})}\BibitemShut {NoStop}%
\end{thebibliography}%

\appendix
\section{Sampling from discrete Wigner functions of spin clusters}\label{app:sampling-discrete-Wigner-cluster}
In this section, we first recapitulate how to derive the concrete sampling rules of dTWA and then extend the scheme to clusters of multiple spins. Finally, we provide a concrete example of sampling rules for clusters of 2 spins. The concepts described here are similar to \cite{Pucci2016} Appendix~A.

\subsection{Recap: Sampling a single spin}
Earlier in section~\ref{sec:discrete-wigner-function}, we defined the phase-point operators for a single spin
\begin{equation}
\hat{A}_{p,q} = (\mathds{1}+\mathbf{r}(p,q) \cdot \mathbf{\hat\sigma})/2    
\end{equation}
via a choice of phase-point vectors $\mathbf{r}(p,q)$. In principle there are many possible choices for $\mathbf{r}(p,q)$ but since there are a total of 8 discrete spin states, two sets of phase point operators are enough to cover all possible states. We define the two sets of phase-point operators (see \cite{Czischek2018} Fig.~2 for a visualization) via
\begin{subequations}
\begin{align}
    r^1(0,0) &= (1,1,1),\\
    r^1(0,1) &= (-1,-1,1),\\
    r^1(1,0) &= (1,-1,-1),\\
    r^1(1,1) &= (-1,1,-1),
\end{align}
\end{subequations}
and
\begin{subequations}
\begin{align}
    r^2(0,0) &= (1,-1,1),\\
    r^2(0,1) &= (-1,1,1),\\
    r^2(1,0) &= (1,1,-1),\\
    r^2(1,1) &= (-1,-1,-1).
\end{align}
\end{subequations}
With this choice, we can define the Wigner function of some quantum state $\hat\rho$ as 
\begin{equation}
    w^s(p,q;\hat\rho) = \frac{1}{2} \operatorname{Tr} \hat\rho \hat{A}^s_{(p,q)}
\end{equation}
where $s=1,2$ denotes the set of phase-point operators. These are normalized for each set, i.e.\ $\sum_{p,q} w^s(p,q;\hat\rho) = \operatorname{Tr}\hat{\rho}=1$ independent of $s$, and thus quasi-probability distributions. In case all values of a $w^s(p,q;\hat\rho)$ are positive, we can treat it as a probability distribution and sample initial conditions for the truncated Wigner approximation from it. For a single spin, it is always possible to rotate the phase-point operators to render the Wigner functions positive, so we can always sample from either one of the two possible choices of Wigner functions. In fact it is crucial to employ both choices for sampling to prevent the introduction of spurious correlations (see example below)~\cite{Czischek2018,minkHybridDiscretecontinuousTruncated2022}. To be explicit, the complete sampling procedure for a single trajectory first randomly selects one of the phase-space representation and then draws a phase-space vector according to its Wigner function.

We illustrate this prescription using the state $\rho=\ket{\uparrow}\bra{\uparrow}$ as an example. The Wigner functions read:
\begin{subequations}
\begin{align}\label{app:eq:single-spin-wigner-1}
    \mathbf{w}^1(\rho) &= \begin{pmatrix} w^1(0,0) & w^1(0,1) \\ w^1(1,0) &w^1(1,1)\end{pmatrix} = \begin{pmatrix}\frac{1}{2} & \frac{1}{2}\\ 0 &0\end{pmatrix}\\
    \mathbf{w}^2(\rho) &= \begin{pmatrix} w^2(0,0) & w^2(0,1) \\ w^2(1,0) &w^2(1,1)\end{pmatrix} = \begin{pmatrix}\frac{1}{2} & \frac{1}{2}\\ 0 &0\end{pmatrix}\label{app:eq:single-spin-wigner-2}
\end{align}
\end{subequations}
Choosing one these Wigner functions at random and then sampling from it is equivalent to drawing a sample from the set $\{r^1(0,0),r^1(0,1),r^2(0,0),r^2(0,1)\}$. In turn, this just means, we need to set the $z$-component to $1$ and choose $x$ and $y$ independently from $\pm 1$. We remark that this prescription reproduces all moments of the spin operators 
$\braket[1]{(\hat\sigma_{x,y,z})^k}$ in contrast to the Gaussian approximation which reproduces means and covariances only. Additionally, we remark that every possible phase-point of $\mathbf{w}^1$ ($\mathbf{w}^2$) has the $x$ and $y$-components aligned (anti-aligned), which is the spurious correlation mentioned earlier. By using both Wigner functions, we avoid artifacts caused by this, making the simulation more accurate.

\subsection{Generalization to clusters of spins}
The prescription, we just outlined, readily generalizes to clusters of spins by taking tensor products of the phase-point operators. Consider a cluster of $n$ spins: The joint Hilbert space is now $SU(D)$, where $D=2^n$, given from the tensor product of Hilbert spaces of the single spins. In the following, we essentially repeat the construction from before applied to the cluster's Hilbert space and exploit its product structure. We denote the operator basis of a cluster of $n$ spins by $\mathbf{X}_n$, which can be constructed recursively by
\begin{subequations}
\begin{align}\label{app:eq:operator-basis-1}
    \mathbf{\hat{X}}_1 &\equiv \mathbf{\hat\sigma}\\\label{app:eq:operator-basis-2}
    [\mathbf{\hat{X}}_n]_{i} &= \begin{cases}
        \hat{\sigma}_i\otimes\mathds{1} & i \in \{1,2,3\}\\
        \mathds{1}\otimes \hat{\sigma}_{i-3} & i \in \{4,5,6\}\\
        [\mathbf{\hat{X}}_1 \otimes  \mathbf{\hat{X}}_{n-1}]_{i-6} & \text{else}
    \end{cases}\quad,
\end{align}
\end{subequations}
where $[\cdot]_i$ denotes the $i$-th component of the vector.

In much the same way, we can construct the phase-point vectors. However, we need to consider that we have two possible choices for each spin to make, so there are a total of $2^n$ sets of phase-point operators. Using $\mathbf{s} \in \{1,2\}^n$, we can construct the phase-point vectors corresponding to the operator basis defined above as:
\begin{align}
    \mathbf{r}^\mathbf{s}_{n}(\mathbf{p},\mathbf{q}) = &\quad\: \mathbf{r}^{s_1}(p_1, q_1) \\\notag
    & \oplus \mathbf{r}^{\mathbf{\tilde s}}_{n-1}(\mathbf{\tilde p},\mathbf{\tilde q})\\\notag
    & \oplus \left[\mathbf{r}^{s_1}(p_1, q_1)\otimes \mathbf{r}^{\mathbf{\tilde s}}_{n-1}(\mathbf{\tilde p},\mathbf{\tilde q})\right]\label{app:eq:phase-space-vector}
\end{align}
where the vectors with tilde ($\mathbf{\tilde s}$, $\mathbf{\tilde p}$ and $\mathbf{\tilde q}$) are the same as the bare vectors without the first element, e.g.\ $\mathbf{\tilde s} = (s_2,\ldots,s_n)$.
From these building blocks, we can define the Wigner functions of the cluster as
\begin{align}
    w^\mathbf{s}(\mathbf{p},\mathbf{q};\hat\rho) &= \frac{1}{2^n} \operatorname{Tr} \hat\rho \hat{A}^\mathbf{s}_{(\mathbf{p},\mathbf{q})}\\
    &= \frac{1}{4^n} \operatorname{Tr} \hat{\rho} \left(\mathds{1} + \mathbf{r}_n^\mathbf{s}(\mathbf{p},\mathbf{q})\cdot \mathbf{\hat X}_n\right).
\end{align}
As can be checked easily via induction, this definition gives us a normalized Wigner function for every choice of $\mathbf{s}$.

Another short calculation shows, that if the quantum state $\hat\rho$ factorizes between the spins, i.e.\ $\rho = \bigotimes_{i\leq n} \hat\rho_i$, then the Wigner function factorizes as well:
\begin{equation}
    w^\mathbf{s}(\mathbf{p},\mathbf{q};\bigotimes_{i\leq n} \hat\rho_i) = \prod_{i\leq n} w^{s_i}(p_i,q_i; \hat\rho_i)
\end{equation}
This allows for efficient sampling. 

To derive rules for sampling initial states, conceptually one needs to choose a random set of phase-point representations, i.e.\ draw $\mathbf{s}$ randomly, and then choose a phase-space vector $\mathbf{r}^\mathbf{s}_n(\mathbf{p},\mathbf{q})$ with a probability determined by the corresponding Wigner function $w^\mathbf{s}(\mathbf{p},\mathbf{q})$. In case of a product initial state, this prescription simplifies dramatically because we choose the phase-space vector of each spin independently and compute the initial value of correlators by products (see Eq.~\ref{app:eq:phase-space-vector}).

We illustrate the prescription given above using the Neel-state $\hat\rho = \ket{\uparrow\downarrow}\bra{\uparrow\downarrow}$. Applying the rule for product states, we can immediately state the sampling scheme: Set $\braket[1]{\hat\sigma_z^1} = -\braket[1]{\hat\sigma_z^2} = 1$, choose $\braket[1]{\hat\sigma_x^1}$,$\braket[1]{\hat\sigma_y^1}$,$\braket[1]{\hat\sigma_x^2}$,$\braket[1]{\hat\sigma_y^2}$ randomly from $\{-1,1\}$ and then compute the initial values of the correlators by products, e.g.\ $\braket[1]{\hat\sigma_x^1\hat{\sigma}_x^2}=\braket[1]{\hat\sigma_x^1}\braket[1]{\hat\sigma_x^2}$.

\begin{table}[t]
    \centering
    \def\arraystretch{1.35}
    \begin{tabular}{c|c|c|c}
        Index $i$ & Operator $X_i$ & Initial value & Term in Eq.~\ref{app:eq:initial-phase-point} \\
        \hline
        1 & $\braket[1]{\hat{\sigma}_x^1}$ & $1$ & \multirow{3}{*}{$\mathbf{r}^1(0,0)$}\\
        2 & $\braket[1]{\hat{\sigma}_y^1}$ & $1$ & \\
        3 & $\braket[1]{\hat{\sigma}_z^1}$ & $1$ & \\
        \hline
        4 & $\braket[1]{\hat{\sigma}_x^2}$ & $1$ & \multirow{3}{*}{$\mathbf{r}^1(1,0)$}\\
        5 & $\braket[1]{\hat{\sigma}_y^2}$ & $1$ & \\
        6 & $\braket[1]{\hat{\sigma}_z^2}$ & $-1$ & \\
        \hline
        7 & $\braket[1]{\hat{\sigma}_x^1\hat{\sigma}_x^2}$ & $1$ & \multirow{9}{*}{$\mathbf{r}^1(0,0)\otimes\mathbf{r}^1(1,0)$}\\
        8 & $\braket[1]{\hat{\sigma}_x^1\hat{\sigma}_y^2}$ & $1$ \\
        9 & $\braket[1]{\hat{\sigma}_x^1\hat{\sigma}_z^2}$ & $-1$ \\
        10 & $\braket[1]{\hat{\sigma}_y^1\hat{\sigma}_x^2}$ & $1$ \\
        11 & $\braket[1]{\hat{\sigma}_y^1\hat{\sigma}_y^2}$ & $1$ \\
        12 & $\braket[1]{\hat{\sigma}_y^1\hat{\sigma}_z^2}$ & $-1$ \\
        13 & $\braket[1]{\hat{\sigma}_z^1\hat{\sigma}_x^2}$ & $1$ \\
        14 & $\braket[1]{\hat{\sigma}_z^1\hat{\sigma}_y^2}$ & $1$ \\
        15 & $\braket[1]{\hat{\sigma}_z^1\hat{\sigma}_z^2}$ & $-1$
    \end{tabular}
    \caption{Coefficients for the phase-point vector given in Eq.~\ref{app:eq:initial-phase-point}}
    \label{app:tab:initial-state}
\end{table}

Alternatively, we can employ the tedious route and compute all the Wigner functions. We start by computing the single spin Wigner functions, which for $\ket\uparrow\bra\downarrow$ are given in Eqs.~\ref{app:eq:single-spin-wigner-1} and \ref{app:eq:single-spin-wigner-2}. Similarly for $\ket\downarrow\bra\downarrow$, we find
\begin{equation}
    \mathbf{w}^1(\ket\downarrow) = \mathbf{w}^2(\ket\downarrow) = \begin{pmatrix}0 & 0\\ \frac{1}{2} &\frac{1}{2}\end{pmatrix}\quad.
\end{equation}
From this we can compute the full two-spin Wigner functions:
\begin{align}\notag
    \mathbf{w}^{(1,1)}&=\mathbf{w}^{(2,2)} = \mathbf{w}^1(\ket{\uparrow})\otimes \mathbf{w}^1(\ket\downarrow)\\\notag
    &= \begin{pmatrix}
        w^1(\ket{\uparrow})(0,0)\cdot \mathbf{w}^1(\ket\downarrow)&
        w^1(\ket{\uparrow})(0,1)\cdot \mathbf{w}^1(\ket\downarrow)& \\
        w^1(\ket{\uparrow})(1,0)\cdot \mathbf{w}^1(\ket\downarrow)& 
        w^1(\ket{\uparrow})(1,1)\cdot \mathbf{w}^1(\ket\downarrow)
    \end{pmatrix}\\
    &= \begin{pmatrix}
        0&0&0&0\\
        \frac{1}{4}&\frac{1}{4}&\frac{1}{4}&\frac{1}{4}\\
        0&0&0&0\\
        0&0&0&0
    \end{pmatrix}\\
    \mathbf{w}^{(1,2)}&=\mathbf{w}^{(2,1)} = \begin{pmatrix}
        0&0&0&0\\
        0&0&0&0\\
        \frac{1}{4}&\frac{1}{4}&\frac{1}{4}&\frac{1}{4}\\
        0&0&0&0
    \end{pmatrix}
\end{align}

To generate a single sample, we first need to select one of the 4 Wigner functions, e.g.\  $w^{(1,1)}$. This Wigner function gives us the probability distribution to choose the state from, which in in this case means, we need to select one of the phase-points $(\mathbf{p},\mathbf{q})$ from the set
\begin{align}
\bigg\{
    &\left[\begin{pmatrix}0\\1\end{pmatrix}, \begin{pmatrix}0\\0\end{pmatrix}\right],
    \left[\begin{pmatrix}0\\1\end{pmatrix}, \begin{pmatrix}0\\1\end{pmatrix}\right],\notag\\
    &\left[\begin{pmatrix}0\\1\end{pmatrix}, \begin{pmatrix}1\\0\end{pmatrix}\right],
    \left[\begin{pmatrix}0\\1\end{pmatrix}, \begin{pmatrix}1\\1\end{pmatrix}\right]
\bigg\}    
\end{align}
with equal probability. Assuming we selected the first phase-point, then the corresponding phase-space vector is given by:
\begin{align}\label{app:eq:initial-phase-point}
    \mathbf{r}^{(1,1)}_2(\begin{pmatrix}0\\1\end{pmatrix},\begin{pmatrix}0\\0\end{pmatrix}) &= \mathbf{r}^1(0,0)\oplus \mathbf{r}^1(1,0) \oplus (\mathbf{r}^1(0,0) \otimes \mathbf{r}^1(1,0))
\end{align}
The corresponding initial values of the trajectory are given explicitly in Tab.~\ref{app:tab:initial-state}.

\section{Single pair dynamics}\label{app:single-pair-dynamics}

\begin{figure}[ht]
    \centering
    \includegraphics[width=1.\columnwidth]{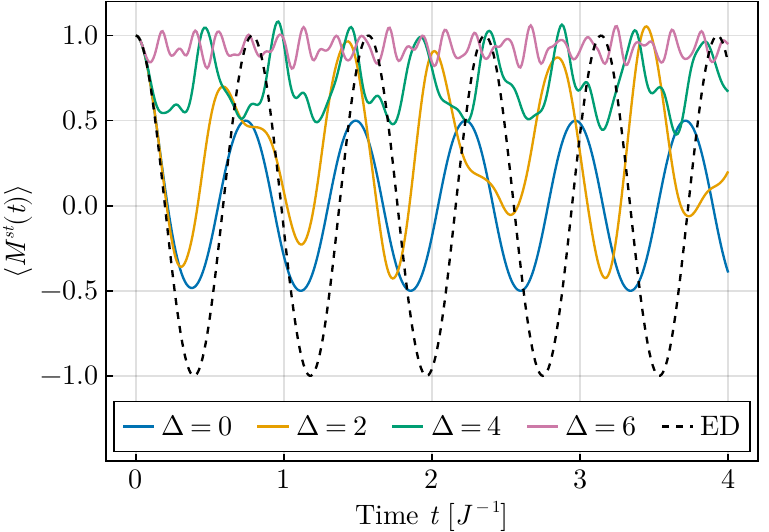}
    \caption{Dynamics of the staggered magnetization for 2 spins with XXZ interaction for various anisotropies $\Delta$. Shown is the exact solution (black, dashed) and solutions obtained with dTWA (colors, solid). The exact dynamics are independent of $\Delta$, so only a single curve is shown.}
    \label{app:fig:two-spin-dynamics}
\end{figure}

To illustrate the inaccuracy of dTWA in the presence of XX interactions, we study the same system as in the main text for two spins. Repeating the definition here for convenience, we consider the Hamiltonian

\begin{equation}
    \hat{H} = 2J\left( \hat\sigma_x^1 \hat\sigma_x^2+ \hat\sigma_y^1 \hat\sigma_y^2 \right) + 2\Delta \hat\sigma_z^1 \hat\sigma_z^2 \quad,
\end{equation}
the initial state $\ket{\psi_0} = \ket{\uparrow\downarrow}$ and the observable $\hat M^{st} = \frac{1}{2} (\hat{\sigma}_z^1+\hat{\sigma}_z^2)$. Since this Hamiltonian conserves total $z$-magnetization $\hat M_z=\hat\sigma_z^1+\hat\sigma_z^2$, the dynamics stays confined to the zero magnetization sector, where the state oscillates back and forth between $\ket{\uparrow\downarrow} \leftrightarrow \ket{\downarrow\uparrow}$. So the exact solution reads $\braket[1]{\hat M^{st}(t)} = \cos(8Jt)$. This is independent of $\Delta$, because the ZZ-term $\hat\sigma_z^1 \hat\sigma_z^2$ of course commutes with $\hat M_z$ and thus cannot introduce additional couplings.

Setting $J=1$ and using dTWA to solve the dynamics for several values of $\Delta$, we see that the semi-classical solution is both influenced strongly by the value of $\Delta$ and yields inaccurate results even for $\Delta=0$ (cf.\ Fig.~\ref{app:fig:two-spin-dynamics}). 

\end{document}